\documentclass{pasa}%

\usepackage{graphicx}
\usepackage{amsmath}	
\usepackage{amssymb}	
\def\degr{$^\circ$}

\title[Rapid ASKAP Continuum Survey]{The Rapid ASKAP Continuum Survey I: Design and First Results}

\author[McConnell et al.]{D.~McConnell$^{1}$, C.~L.~Hale$^{2}$, E.~Lenc$^{1}$, 
J.~K.~Banfield$^{1}$, George~Heald$^{2}$, A.W.~Hotan$^{2}$, James~K.~Leung$^{3,1}$, 
Vanessa~A.~Moss$^{1,3}$, Tara~Murphy$^{3}$, Andrew~O'Brien$^{1,4,5}$, Joshua~Pritchard$^{1,3}$, 
Wasim~Raja$^{1}$, Elaine~M.~Sadler$^{1,3}$, Adam~Stewart$^{3}$, Alec~J.~M.~Thomson$^{2}$, 
M.~Whiting$^{1}$, James~R.~Allison$^{6}$, S.W.~Amy$^{1}$, C.~Anderson$^{2,7}$, 
Lewis~Ball$^{1,8}$, Keith~W.~Bannister$^{1}$, Martin~Bell$^{1,9}$, Douglas~C.-J.~Bock$^{1}$, 
Russ~Bolton$^{1}$, J.~D.~Bunton$^{1}$, A.~P.~Chippendale$^{1}$, J.~D.~Collier$^{1,10,5}$, 
F.~R.~Cooray$^{1}$, T.J.~Cornwell$^{1,11}$, P.J.~Diamond$^{1,8}$, P.~G.~Edwards$^{1}$, 
N.~Gupta$^{1,12}$, Douglas~B.~Hayman$^{1}$, Ian~Heywood$^{6,13}$, C.~A.~Jackson$^{1,14}$, 
B\"arbel~S.~Koribalski$^{1,5}$, Karen~Lee-Waddell$^{1}$, N.~M.~McClure-Griffiths$^{15}$, Alan~Ng$^{1}$, 
Ray~P.~Norris$^{1,5}$, Chris~Phillips$^{1}$, John~E.~Reynolds$^{1}$, Daniel~N.~Roxby$^{1}$, 
Antony~E.T.~Schinckel$^{1}$, Matt~Shields$^{1}$, Chenoa~Tremblay$^{2}$, A.~Tzioumis$^{1}$, 
M.A.~Voronkov$^{1}$, Tobias~Westmeier$^{16}$
\affil{$^{1}$CSIRO Astronomy and Space Science, PO Box 76 Epping NSW 1710 Australia}
\affil{$^{2}$CSIRO Astronomy and Space Science, PO Box 1130, Bentley WA 6102, Australia}
\affil{$^{3}$Sydney Institute for Astronomy, School of Physics, University of Sydney, NSW 2006, Australia}
\affil{$^{4}$Center for Gravitation, Cosmology, and Astrophysics, Department of Physics, University of Wisconsin-Milwaukee, Milwaukee, WI 53201, USA}
\affil{$^{5}$Western Sydney University, Locked Bag 1797, Penrith, NSW 2751, Australia}
\affil{$^{6}$Sub-dept of Astrophysics, Physics, University of Oxford, Denys Wilkinson Building, Keble Road, Oxford OX1 3RH, UK}
\affil{$^{7}$Jansky fellow of the National Radio Astronomy Observatory, NRAO, 1003 Lopezville Rd, Socorro, NM 87801 USA}
\affil{$^{8}$SKA Organisation, Jodrell Bank, Lower Withington, Macclesfield, Cheshire SK11 9FT, UK}
\affil{$^{9}$University of Technology Sydney, 15 Broadway, Ultimo NSW 2007}
\affil{$^{10}$The Inter-University Institute for Data Intensive Astronomy (IDIA), Department of Astronomy, University of Cape Town, Private Bag X3, Rondebosch, 7701, South Africa }
\affil{$^{11}$Tim Cornwell Consulting, 17 Elgan Crescent, Sandbach, CW111LD, UK}
\affil{$^{12}$Inter-University Centre for Astronomy and Astrophysics, Post Bag 4, Ganeshkhind, 411007 Pune, India}
\affil{$^{13}$Department of Physics \& Electronics, Rhodes University, Makhanda, 6139, South Africa}
\affil{$^{14}$ASTRON, The Netherlands Institute for Radio Astronomy, Oude Hoogeveensdijk 4, Dwingeloo, 7991 PD, NL}
\affil{$^{15}$Research School of Astronomy \& Astrophysics, Australian National University,  Canberra, Australia 2611}
\affil{$^{16}$ICRAR, M468, The University of Western Australia, 35 Stirling Hwy, Crawley WA 6009, Australia}
}
\jid{PASA}
\doi{ https://doi.org/10.1017/pasa.2020.41}
\jyear{\the\year}

\usepackage{aas_macros}
\usepackage{hyperref} 
\hypersetup{colorlinks,citecolor=blue,linkcolor=blue,urlcolor=blue}

\begin{document}

\begin{frontmatter}
\maketitle

\begin{abstract}
The Rapid ASKAP Continuum Survey (RACS) is the first large-area survey to be conducted with the full 36-antenna Australian Square Kilometre Array Pathfinder (ASKAP) telescope.
RACS will provide a shallow model of the ASKAP sky that will aid the calibration of future deep ASKAP surveys.
RACS will cover the whole sky visible from the ASKAP site in Western Australia, and will cover the full ASKAP band of $700-1800$~MHz. The RACS images are generally deeper than the existing NRAO VLA Sky Survey (NVSS) and Sydney University Molonglo Sky Survey (SUMSS) radio surveys and have better spatial resolution. 
All RACS survey products will be public, including radio images (with $\sim 15$ arcsecond resolution) and catalogues of about three million source components with spectral index and polarisation information. In this paper, we present a description of the RACS survey and the  first data release of 903 images covering the sky south of declination $+41^\circ$ made over a 288\,MHz band centred at 887.5\,MHz.
\end{abstract}

\end{frontmatter}

\section{INTRODUCTION }
\label{sec:intro}
The Rapid ASKAP Continuum Survey (RACS) is a shallow all-sky precursor to the full, multi-year surveys to be conducted with the Australian SKA\footnote{Square Kilometre Array} Pathfinder \citep[ASKAP;][]{Johnston:2007ku, Hotan:2020}. It will image the entire sky south of declination $\delta = +51$\degr\ over representative bands within ASKAP's operating frequency range of 700--1800\,MHz. The aims of this survey are to generate reference images to aid ASKAP's future operation, to exercise the newly commissioned instrument ahead of the commencement of its full scientific operations, and to provide a valuable astronomical resource. 

ASKAP was designed to be a survey instrument capable of quickly observing the whole accessible sky. It is located at the Murchison Radio-astronomy Observatory (MRO) in Western Australia, and is operated by the Commonwealth Scientific and Industrial Research Organisation (CSIRO). ASKAP is an array of 36 12-metre prime-focus antennas; each is equipped with a phased array feed (PAF) that enables the simultaneous digital formation of 36 dual-polarisation beams to sample its 31 square degree field of view. It has an instantaneous bandwidth of 288\,MHz. A full description of ASKAP is in preparation \citep{Hotan:2020}, but descriptions of the PAFs, beam formation, and telescope operation exist in \citet{Hotan:2014dv} and \citet{McConnell:2016fu}.

The data gathering capacity of ASKAP, equal to 36 simultaneous 630-baseline synthesis arrays, presented a software development challenge: how to develop calibration and imaging software that could run in less time than the time taken to make the observations. In addition to the use of highly parallel super-computers, the solution relied upon the availability of a model of the sky that could provide the properties of all the major sources in any field being observed \citep{Cornwell:2011vq}. The aim was to construct this model (the global sky model---GSM) from short observations made with ASKAP itself, and that all subsequent observations would contribute to its improvement. Although the early operation of ASKAP makes no attempt to form images in real-time, the availability of a sky model will assist data calibration and reduce reliance on time-consuming observations of calibration sources. The survey we describe here will allow the initialisation of the Global Sky Model. 


RACS will be a valuable resource and will complement other all-sky radio surveys.
Table \ref{tab:surveys} gives a comparison of the RACS survey parameters with other comparable radio surveys with substantial Southern-sky coverage in the metric and decimetric bands: VLSSr \citep[Very Large Array Low-frequency Sky Survey Redux;][]{lane_etal_2014}, GLEAM \citep[Galactic and Extragalactic All-sky Murchison Widefield Array survey;][]{wayth_etal_2015, 2017MNRAS.464.1146H, 2018MNRAS.478.2835L}, TGSS-ADR1 \citep[TIFR GMRT Sky Survey;][]{Intema:2017bd}, NVSS \citep[NRAO VLA Sky Survey;][]{1998AJ....115.1693C}, SUMSS \citep[Sydney University Molonglo Sky Survey;][]{Mauch:2003ed}, MGPS-2 \citep[Molonglo Galactic Plane Survey;][]{Murphy:2007kx}, and the VLASS \citep[VLA Sky Survey;][]{lacy_etal_2020}.  Together, NVSS in the north and SUMSS and MGPS-2 in the south cover the whole sky and have been the primary reference for radio sources at gigahertz frequencies. It is clear from Table~\ref{tab:surveys} that RACS fills a critical niche connecting low-frequency surveys at metre wavelength to existing and forthcoming decimetric surveys. Sensitive wideband coverage in this intermediate regime strengthens efforts to understand the broadband spectra of the radio source population \citep[e.g.,][]{callingham_etal_2017,degasperin_etal_2018}. RACS also establishes a solid reference catalogue against which to assess radio source variability and transient candidates as future surveys emerge from ASKAP, MeerKAT and the SKA.

\begin{table*}
\caption{Summary of RACS parameters with those of other comparable surveys. The tabulated data allow comparison with RACS; for detailed information consult the reference papers mentioned in Section \ref{sec:intro}.}
\centering
\begin{tabular}{lccccccc}
\hline\hline
Survey & Frequency & Bandwidth & Resolution & Sky coverage & Sensitivity & Polarization & N$_\mathrm{sources}$ \\
& (MHz) & (MHz) & (arcsec) & (sq deg) & (mJy/beam) &  & ($\times10^6$) \\
\hline%
VLSSr & 73.8 & 3.12 & 75 & 30,793 & 100 & I & 0.93 \\
GLEAM & 87, 118, 154, & 30.72 & 120 & 27,691 & 6--10 & I,Q,U,V & 0.33 \\
& 185, 215 & & & & & \\
TGSS & 150 & 16.7& 25 & 36,900 & 2--5 & I & 0.62 \\
RACS\textsuperscript{1} & 887.5 & 288 & 15 & 36,656 & $\sim$0.25 & I,Q,U,V & 4 \\
& 1295.5 &  & & & & & \\
& 1655.5 &  & & & & & \\
RACS\textsuperscript{2} & 887.5 & 288 & 15--25 & 34,240 & 0.2--0.4 & I & 2.8 \\
SUMSS & 843 & 3 & 45 & 10,300 & 1.5 & RC & 0.2 \\
+MGPS-2 & & & & & & \\
NVSS & 1346, 1435 & 42 & 45 & 33,800 & 0.45 & I,Q,U & 2 \\
VLASS & 3000 & 2000 & 2.5 & 33,885 & 0.07 & I,Q,U & 5.3 \\
\hline\hline
\multicolumn{7}{l}{\textsuperscript{1}\footnotesize{RACS full survey capability.}} \\
\multicolumn{7}{l}{\textsuperscript{2}\footnotesize{RACS first data release.}} \\
\end{tabular}
\label{tab:surveys}
\end{table*}

This paper is a companion to the first RACS data release of 903 images made at a centre frequency of 887.5\,MHz and covering the sky south of $\delta = +41$\degr (83 per cent of the celestial sphere). We outline the survey design in Section \ref{sec:plan}, and how it uses the established capabilities of the telescope.  In  Section \ref{sec:survey_results} we describe the specific approach taken with the first epoch of observations and present their results. We discuss image quality and describe some extra steps taken to optimise the scientific utility of the RACS data products. Section \ref{sec:datarelease} lists the products to be released, and gives some examples of RACS images.  In Section \ref{sec:summary} we summarise the the current state of the survey and our plans for its future.

\section{Survey Design}
\label{sec:plan}
RACS observations cover the whole accessible sky over representative portions of ASKAP's operating frequency range. Each observation is made with the telescope's full instantaneous bandwidth of 288\,MHz, divided into 288 contiguous frequency channels.  To reach the RACS point-source detection sensitivity goal of approximately 1\,mJy an integration time of  $\sim$1000\,s per pointing is needed.
All four polarisation products (XX, XY, YX, YY) are recorded to allow images to be made in Stokes parameters $I$, $Q$, $U$, and $V$. The data are processed on the facilities of the Pawsey Supercomputing Centre using the ASKAPsoft software package \citep{Cornwell:2011}, and images are made available on the CSIRO ASKAP Science Data Archive---CASDA\footnote{\url{ https://research.csiro.au/casda/}} \citep[CASDA;][]{Chapman:2017}.

A catalogue of total intensity source components identified in RACS images is being constructed and will be described in a second paper, ``paper II''  (Hale et al. {\em in prep}) and will be released on CASDA.  The calibrated RACS data are also being used to prepare frequency-resolved images in linear polarisation that will yield a catalogue of rotation-measures across the survey area (Thomson  et al., {\em in prep}).

\subsection{Spectral coverage}

\begin{figure}
	\includegraphics[width=\columnwidth]{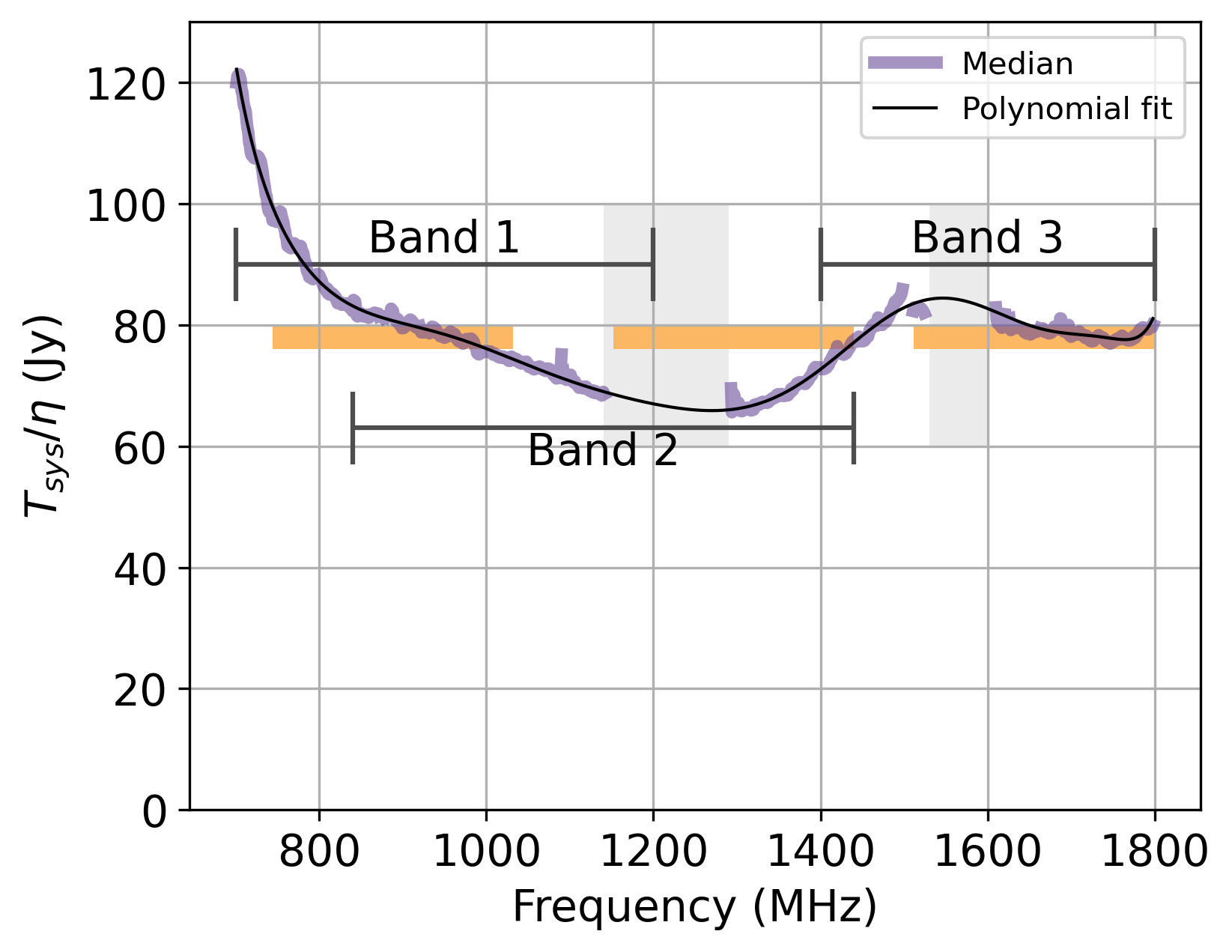}
    \caption{ASKAP sensitivity over its frequency range. The purple line traces the median value of central beams of all 36 antennas.  Bands affected by radio-frequency interference are shaded grey.  ASKAP's three tuning ranges are shown and labelled. The proposed RACS bands are shown as orange bars.}
    \label{fig:frequency_coverage}
\end{figure}

The full spectral coverage of ASKAP spans 700--1800~MHz, however each observation is restricted to a contiguous 288\,MHz band within that range. The choice of observing band has constraints based on both hardware and the presence of airborne and orbiting sources of radio-frequency interference (RFI). Figure \ref{fig:frequency_coverage} shows the proposed RACS frequency coverage. The first RACS data were collected over the 288~MHz wide band centred at 887.5~MHz. Subsequent RACS observations will be tuned to centre frequencies of 1295.5 and 1655.5\,MHz; however, parts of these bands will be severely affected by RFI.

\subsection{Field of view}
ASKAP's field-of-view is approximately square.
Sensitivity measurements across the field-of-view show that its shape is in good agreement with the predictions of modelling by \citet{Bunton:2010vs}, and that it has an equivalent area of 31 square degrees, independent of frequency. Note that the full field-of-view cannot be sampled at the shorter wavelengths with the 36 beams available \citep{Hotan:2020}.

Beams are formed to lie within the field-of-view, giving sensitivity to sky emission over a tile\footnote{We use three terms for similar, but not identical concepts: a ``mosaic'' is any image formed by linear combination of another set of (usually) overlapping images; a ``footprint'' is the arrangement and distribution of electronically formed beams within the field-of-view; and a ``tile'' is the patch of sky imaged in a single ASKAP pointing.} of area $\leq$31 square degrees (Figure \ref{fig:single_tile}). The size of each tile is determined by the beam pattern (footprint) and beam spacing. The choice of beam spacing is driven by several factors and depends on the scientific goals. 

Survey speed varies in proportion to
\begin{equation}
\label{eqn:speed}
\int_{\Theta} \frac{1}{\sigma^2(\theta,\phi)} d\theta d\phi
\end{equation}
where $\sigma^2(\theta,\phi)$ is the system noise variance over the tile; $\theta$, $\phi$ are orthogonal angular coordinates over the solid angle $\Theta$ subtended by the tile.  Random noise is partially correlated between adjacent beams (because of their shared use of adjacent PAF elements), so there is benefit to maximising beam spacing.  Survey speed increases with beam spacing until the outer beams encounter the sensitivity drop at the edge of the field-of-view.  On the other hand, minimal sensitivity ripple across the tile and freedom from polarisation leakage far from beam centres is achieved by decreasing the beam spacing.

Images are produced for each tile as the linear mosaic of the images made from each beam.  Each beam image is made over an area that, for most of the frequency range, covers the main lobe of that beam; that is out to or beyond the beam pattern's first null. This provides significant overlap between beam images.

ASKAP uses two geometries in its beam footprints: square and hexagonal\footnote{These are referred to as {\tt square\_6x6} and {\tt closepack36}}.  The square footprint provides a better match to the shape of the field-of-view, whereas the hexagonal pattern has lower sensitivity ripple and polarimetric aberration.
For the first epoch of RACS observations, beams were arranged in a $6\times6$ square grid with a 1.05 degree centre-to-centre separation (pitch) as shown in Figure \ref{fig:single_tile}. At the highest frequency in the band (1031.5\,MHz) the maximum beam spacing (along the grid diagonal) is equal to the beam full width at half maximum (FWHM). See Section \ref{subsubsec:pos_var} for estimates of true beam shapes across this footprint.

\subsection{Sky coverage}
\begin{figure}
	\includegraphics[width=\columnwidth]{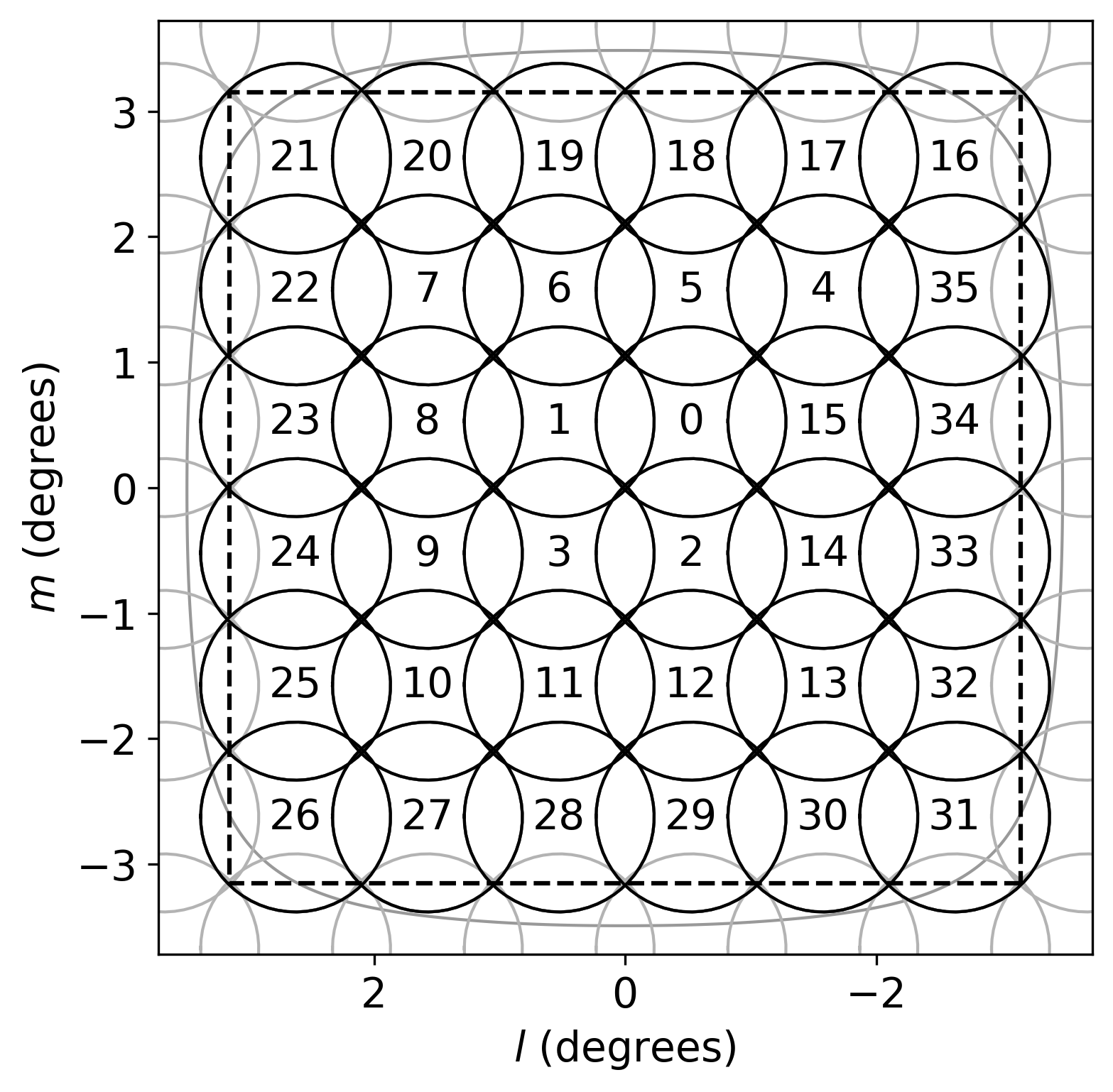}
    \caption{ASKAP field-of-view $(l,m)$ using the {\tt square\_6x6} beam footprint. The positions of the 36 beams (numbered 0--35) are shown as idealised circles at their contour of half-power at 1031~MHz.  In practice the total-intensity beams are close to circular at field centre, but become increasingly elliptical towards the edge of the field with typical eccentricity of $e \lesssim 0.4$. The PAF sensitivity is greatest at the field centre, varies slowly over most of the field, and declines steeply at the edges; the outer grey contour shows the estimated locus of 50\% sensitivity. Beams of adjacent non-overlapping tiles are also shown in grey.}
    \label{fig:single_tile}
\end{figure}

 To sample the sky with uniform sensitivity we construct a tiling of the celestial sphere with square tiles sized to match the beam footprint.  Adjacent non-overlapping tiles are placed so that beam separation across tile boundaries is unchanged.  Figure \ref{fig:single_tile} shows beams of adjacent tiles for the non-overlapping case.
 
\begin{figure}
	\includegraphics[width=\columnwidth]{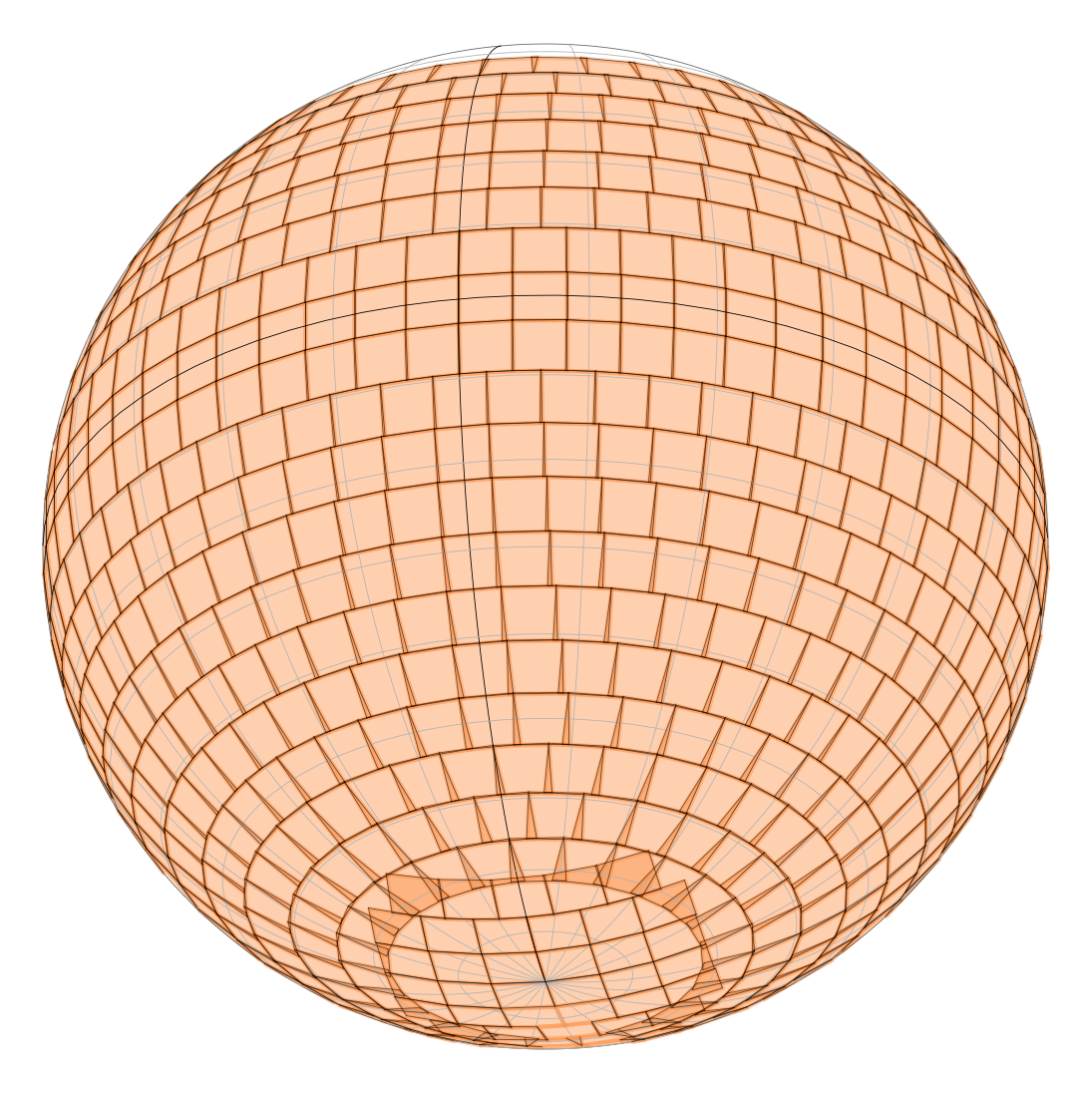}
    \caption{An orthographic view of the celestial sphere showing the arrangement of the RACS observing tiles. Ranks of tiles are centred on a series of declinations from +37.6 to -68.7 degrees, giving full sensitivity from $-$71.3 to +40.2 degrees. A quasi-rectangular grid of tiles is placed over the zone south of declination $-$71 degrees, centred on the south celestial pole.}
    \label{fig:sky_coverage}
\end{figure}

 Tiles are arranged in rows along lines of constant declination; the number of tiles in each row is the minimum needed to span 360 degrees of right ascension without gaps; successive rows are spaced in declination using the same criterion of no gaps but minimal overlap.  Within about 20 degrees of the pole this pattern is replaced by a quasi-rectangular grid centred on the pole. Figure \ref{fig:sky_coverage} illustrates this scheme.

 The absolute northern limit of telescope pointing, $\delta = +48$\degr, is set by the observatory's latitude (26.7 degrees South) and the 15 degree mechanical elevation limit of the ASKAP antennas. ASKAP is sensitive to the sky up to 3 degrees from its pointing direction, and so RACS will cover the sky from the south celestial pole to a declination of $\delta = +51$\degr (although the data presented here are south of $\delta = +41$\degr). 

Overlap between tiles is unavoidable, and varies with declination.  Averaged over the sphere, about six per cent of the surveyed area appears within the full-sensitivity bounds of more than one tile.  This fraction is only mildly dependent on tile size (within the range useful with ASKAP) and on the chosen boundaries of the polar cap.
As mentioned above, even for minimal overlap between tiles, the beam spacing is nowhere greater than it is within each tile, so that image sensitivity is maintained over tile boundaries.  In regions of overlap the sensitivity is improved.

Many sources will be detected in more than one tile, and many more will be detected in several beams.  These multiple detections will be used in the analysis of beam and tile-specific systematic errors, and the tile-tile overlaps will give some visibility of variability on the radio sky.

\subsection{Sampling the angular-scale spectrum}
\label{subsec:uvsampling}
The ASKAP antenna configuration was designed to maximise the sensitivity for extragalactic {\sc Hi} surveys \citep{Gupta:2008vc}, but with additional elements added to the array centre and periphery to improve both the surface-brightness sensitivity and resolution.  The resulting configuration gives a good sensitivity on scales from 10 arcseconds to 3 arcminutes (at 1.4\,GHz, and with suitable visibility weighting). The distribution of samples of the $(u,v)$ plane enabled by ASKAP's configuration is shown in Figure \ref{fig:u-v_sampling}, scaled appropriately for this first set of RACS observations.

Without the benefit of Earth rotation, the inner part of the $(u, v)$ plane is poorly sampled, making it harder to image larger angular scales with the short integrations used by RACS.  The upper right panel of Figure \ref{fig:u-v_sampling} illustrates this limitation, which is most acute for spatial scales larger than ten arcminutes at 887.5~MHz (baselines shorter than 100 metres).

\subsection{The point-spread-function}
\label{subsec:psf}
\begin{figure}
	\includegraphics[width=\columnwidth]{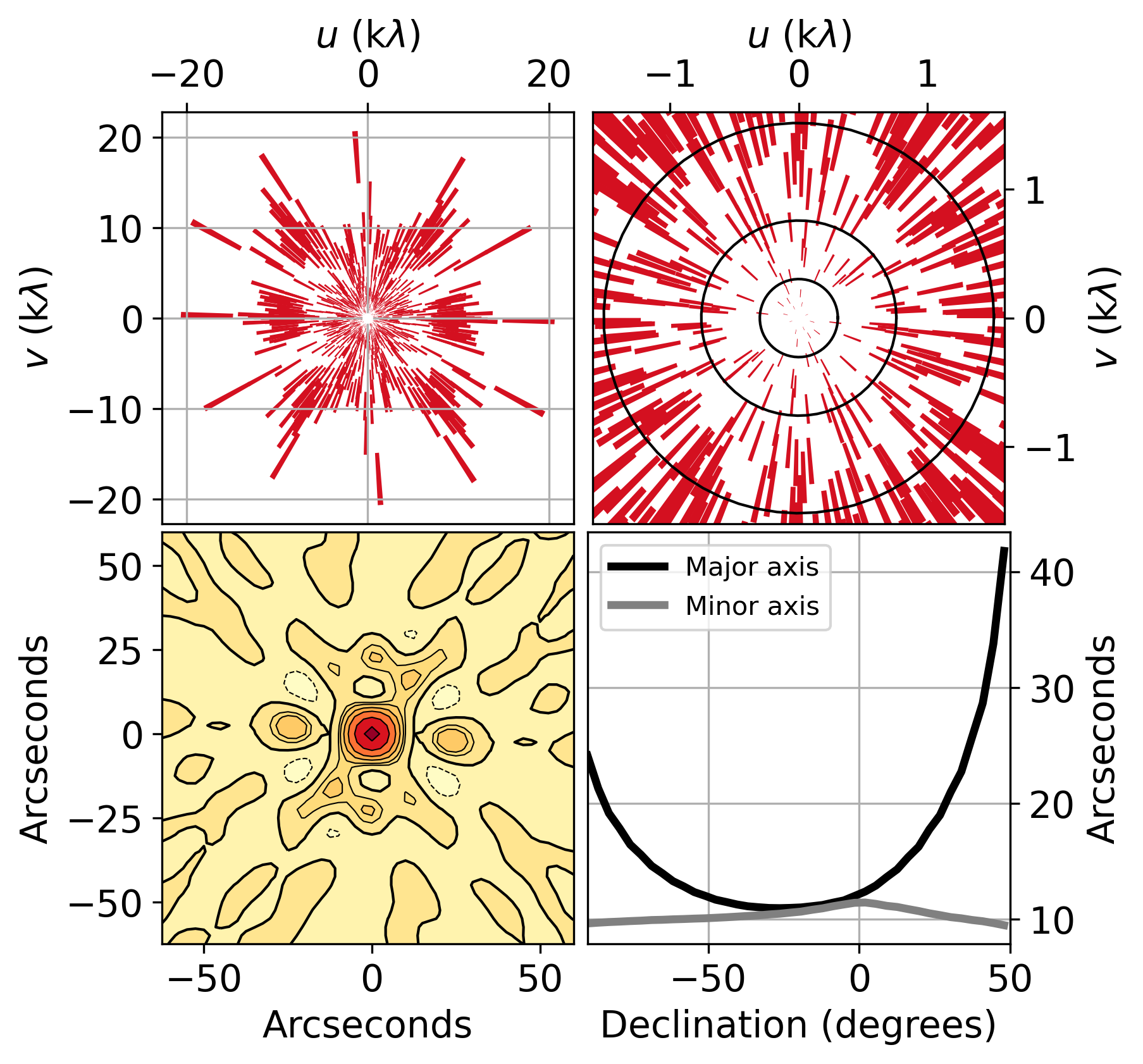}
    \caption{Sampling of $(u,v)$ coordinates over a 15-minute integration for a source near the zenith and the resultant point-spread-function (computed for the observing frequency of the first RACS observations). Upper left: whole $(u,v)$ plane sampled. Upper right: inner part, showing that for extended structures the $(u,v)$ sampling becomes sparse.  The circles correspond to spatial scales of approximately 2 (outer), 4 and 10 (inner) arcminutes for observations made at 887\,MHz.  Lower left: The PSF; the contours lie at 0,  $\pm$10,  15,  24,  37,  60 and  90 per cent of the peak.  Lower right: As a function of declination, the major and minor axis lengths of the main lobe of the PSF using this  $(u,v)$ sampling for observations made on the meridian.}
    \label{fig:u-v_sampling}
\end{figure}

The point-spread-function (PSF) depends on the distribution of projected baseline vectors and on the weighting scheme used for image formation.  For the short integrations ($\sim$1000\,s) used for RACS, the best resolution and PSF symmetry is obtained by observing on the meridian.  Figure \ref{fig:u-v_sampling} (right, lower) shows the PSF major and minor axes (full major and minor axes of an ellipse at the half-height of the PSF) as a function of field declination, assuming a weighting scheme parameterised by robustness $r = 0.0$ ($-2 < r < 2$ : \cite{briggs1995}).

The RACS tile images are composed from 36 beam images, each having different sampling of the $(u, v)$ plane, and so different PSFs.  Even for the ideal case of an observation on the meridian with no data flagging so that all beams use the same set of antennas, the variation in PSF size over the field-of-view exceeds 15 per cent at extreme northern and southern declinations (see Figure \ref{fig:u-v_sampling} - lower right).  This variation increases for observations away from the meridian and in the realistic case of data from some antennas being unusable for some beams.

\subsection{Polarimetry}\label{sec:pol}

A detailed description of the RACS polarisation performance and calibration will be presented elsewhere (Thomson  et al., {\em in prep}). In summary, RACS beams are formed in paired (i.e. co-located on the sky) linear X and Y polarisation, with zero relative phase by construction \citep{Chippendale:2019}. ASKAP's roll axis is driven to maintain the position and orientation of these beams and their planes of polarisation on the sky, simplifying polarimetric calibration and imaging (described further in Section \ref{subsec:calibration}). Thus, Stokes $I$, $Q$, $U$, and $V$ can be reconstructed for each 1 MHz frequency channel in each beam. 

\subsection{Calibration}
\label{subsec:calibration}

\subsubsection{Beam forming}
The RACS observations use the standard ASKAP calibration procedures. Each of the 36 beams are formed with weights that maximise the signal-to-noise ratio in the direction of a strong point source; the Sun is used \citep{Hotan:2020}.  These weights accommodate the different gains of the PAF elements.  Each antenna has an ODC---an ``on-dish-calibrator'' system that radiates a signal from the vertex of the primary reflector and records the response from each PAF element \citep{Chippendale:2019}.  This is used to track changes in PAF element gain and allows beam weights to be adjusted in compensation. Also, the ODC provides a polarised reference signal that is used to align the phase of each dual-polarised beam, that is to set the so-called ``XY--phase'' to zero.

\subsubsection{Antenna gains}
\label{subsubsec:antenna_gains}
The primary calibration of the various aspects of antenna gain is done with an observation of PKS\,B1934--638, the primary cosmic reference source for ASKAP. The term ``antenna gain'' normally refers to those aspects of the synthesis telescope's response that can be factored into antenna-specific quantities.  For ASKAP we use that term for each beam formed on an antenna.  With its 36 formed beams, ASKAP is effectively 36 parallel synthesis telescopes for which each beam is the equivalent of an antenna in a conventional telescope.

For each science observation, ASKAP observes PKS\,B1934--638 with each beam in turn.  The pointing direction and roll axis of the antennas are adjusted so that the reference source lies in the target beam centre and that the instrumental polarisation direction has the same relation to celestial coordinates for all beams. These data are used for:
\begin{description}
\item[Band pass calibration and flux-density scale:] The known spectral energy distribution of PKS\,B1934--638 \citep{Reynolds:1994vd} is used to calibrate both the instrumental band-pass shape and the absolute flux-density scale (but see Section \ref{subsubsec:fluxscale} below for more details).

\item[Interferometer phase:] ASKAP maintains a separate phase and delay tracking reference position in the centre of each beam. Antenna gains are set so that visibility phases on all baselines are zero in the PKS\,B1934--638 data, fixing the position reference for images of the science field. Astrometric accuracy in the science field images may be limited by electronic drifts in the signal path and by differences in path through the atmosphere between the calibration and the science observations.

\item[On-axis polarisation leakage:] At decimetre wavelengths PKS\,B1934--638 is less than 0.1 per cent polarised \citep{Schnitzeler2011} and the polarised emission from nearby sources is negligible. Any observed polarisation therefore arises from coupling between the $X$ and $Y$ beams, and is used to determine coefficients in the feed-error matrix $\mathbf{D}$ \citep{Hamaker:1996wh}, the so-called leakage `D-terms'. For ASKAP, these are found to be low and stable and dominated by a real-valued offset (corresponding to non-orthogonality for the $X$ and $Y$ dipoles) with a magnitude of up to one per cent, but typically less than $0.5$ per cent \citep{Sault2014,Anderson2018}.
\end{description}

\subsubsection{Wide-field instrumental polarisation}
ASKAP's formed beams are similar across antennas, and maintain their position and orientation on the sky (Section \ref{sec:pol}). Thus the pattern of instrumental polarisation over each beam should be stable, and so amenable to being characterised, leading to an image-based correction. \cite{Sault:2015ua} has described such a process, but points out that it will be complicated by any significant time-variation of ionospheric Faraday Rotation.  The development of a wide-field polarisation calibration procedure is required before the release of RACS polarisation products.

\section{Early survey results}
\label{sec:survey_results}
\subsection{Observations}
\begin{table}
\caption{First epoch observation parameters}
\centering
\begin{tabular}{lc}
\hline\hline
Frequency & 887.5\,MHz \\
Bandwidth &  288\,MHz \\
Sky coverage & $-$90\degr < $\delta$ < $+$41\degr \\
Tiling & 903 tiles (see Fig. \ref{fig:sky_coverage})\\
Integration per tile & 15 minutes\\
Footprint & {\tt square\_6x6} \\
Beam spacing & 1.05 deg\\
Surveyed area & 34,240 deg$^2$\\
\hline\hline
\end{tabular}
\label{tab:survey-parameters}
\end{table}

In this paper we present results from the first complete epoch of RACS. The first all-sky observations began on 2019APR21 using the parameters given in Table \ref{tab:survey-parameters}. Observations were scheduled automatically, sequencing each field optimally given a number of constraints that included elevation and distance from solar system bodies. The optimisation took account of the relative availability of fields, mainly determined by time above the horizon, and time-consuming telescope operations such as the rotation of the roll axis. 

Initial sessions in April and May omitted the region around the south celestial pole and fields within about five degrees of the Sun.  The southern tiles were observed in 2019 August, and tiles originally omitted around the Sun, and some others affected by the Moon or instrumental failure were observed in 2019OCT--NOV.
As mentioned in Section \ref{subsubsec:psf} below, many of the early observations were made at a large angular distance from the meridian. The scheduler was improved by adding an hour-angle constraint and the most-affected fields were then re-observed in 2020MAR--JUN. Table \ref{tab:observations} lists the dates of all the observations.

PKS\,B1934$-$638 was observed for 200\,s with each beam, typically once per day. Data from these observations were used for calibration as described in \ref{subsubsec:antenna_gains}. Experience with ASKAP has shown that antenna gains are stable enough to allow calibration of science data up to about 24\,hours after the calibration is taken.  Short term (minute-scale) gain fluctuations are small and can be corrected with self-calibration as described in Section \ref{subsec:calibration-and-imaging}.  Longer term gain amplitude variations are quantified as part of the flux-density calibration described in Section \ref{subsec:fluxscalecorr}. 

\begin{table}
\caption{First epoch observation dates}
\centering
\begin{tabular}{lr}
\hline\hline
Dates & Fields\\ \hline
2019 & \\
\hspace{1em}April 21, 22, 24 -- 30 & 692\\
\hspace{1em}May 4, 6, 7 & 177\\
\hspace{1em}August 3, 17 & 31\\
\hspace{1em}October 31 & 13 \\
\hspace{1em}November 8 & 8\\
2020 & \\
\hspace{1em}March 26 -- 29 & 123 \\
\hspace{1em}April 30 -- May 4 & 168 \\
\hspace{1em}June 19 -- 21 & 24 \\
\hline\hline
\end{tabular}
\label{tab:observations}
\end{table}

\subsection{Calibration and Imaging}
\label{subsec:calibration-and-imaging}
RACS data were processed by the {\sc ASKAPsoft} package \citep{Cornwell:2011, 2019ascl.soft12003G} using the {\em Galaxy} computer cluster that is maintained at the Pawsey Supercomputing Centre. The software is organised into a pipeline---the sequence of {\sc ASKAPsoft} applications used to select, flag, calibrate and image the data.  The pipeline software accepted parameters for controlling how each step is conducted, and the whole system is organised so that all RACS fields were processed consistently.

The following steps were taken:
\begin{itemize}
    \item {\bf Selection : } visibility data from both the calibrator and science fields were copied from the raw MeasurementSets; 
    \item {\bf Flagging : } anomalous samples in the data from both calibrator and science fields were identified using both fixed and dynamically set thresholds in amplitude, and excluded from further use; data judged to be poor following calibration steps were also excluded;
    
    \item {\bf Primary calibration : } data from the observations of PKS\,B1934$-$638 were used to determine the frequency-dependent gains for each beam as described in \ref{subsubsec:antenna_gains}, establishing the instrumental flux-density scale and interferometer phase for each beam centre, as well as factors to correct the on-axis polarisation leakage.
    
    \item {\bf Imaging (total intensity) : } calibrated multi-frequency visibility data for each beam were gridded and imaged as the first two terms (0, 1) of a Taylor series using multi-frequency synthesis. Self-calibration was performed: each beam image yielded a field model that was used to derive  a set of antenna-gain adjustments. For RACS data this cycle was run twice; that is each beam was imaged three times, the second and third after a gain adjustment derived from the previous image. (In future the sky model derived from RACS may be used in this process.)
    
    Images for each beam were made with 6144$\times$6144 cells of size 2.5~arcseconds. This is the largest image size practical using the {\em Galaxy} computer cluster in its current configuration. Each beam image has a 4.2\degr\ extent, which covers the main lobe of the primary beam at the high-frequency end of the band, and reaches the 4 percent sensitivity level at the low-frequency end (see Figure \ref{fig:holo_beams}). The weighting of visibility measurements is achieved by {\em preconditioning} \citep{rau2010}, an approach designed to minimise data access load on the software. RACS imaging used preconditioning that achieved the equivalent of a \cite{briggs1995} ``robust weight'' of $r = 0.0$ ($r$ lies in [-2,2]). The Briggs weighting scheme provides a continuum between the extremes of `natural' (gridded data weighted by the number of measured points) and `uniform' (all gridded data given the same weight). The value $r = 0.0$ gives a compromise between sensitivity, maximised with natural weights, and resolution, better with uniform weights.
    
    Deconvolution used the {\sc BasisfunctionMFS} algorithm\footnote{See \url{https://www.atnf.csiro.au/computing/software/askapsoft/sdp/docs/current/pipelines/introduction.html}}, which supports multi-scale cleaning. It was done in many ``minor cycles'', in which an approximation of the PSF centred on the maximum residual at each iteration was subtracted from the image, and fewer ``major cycles'' in which the modelled sources were subtracted from the visibility data. For the RACS imaging, major cycles used a W-projection algorithm \citep{Cornwell:2008bi} to correct for the effects of non-coplanar baselines with 512 W-planes out to a maximum of 26,000 wavelengths. Each major cycle used a target peak residual of 1\,mJy/beam. The minor cycles used residual thresholds that decreased with each self-calibration cycle:  100, 10 and  1\,mJy/beam.  In the final imaging cycle we used three cleaning scales of 0, 20 and 50 pixels.
    
    Full complex gain corrections (amplitude and phase) were determined in the self calibration step, but the solutions were normalised before use to approximate the traditional ``phase-only'' self calibration. Visibility data from short baselines (< 200~metres, corresponding to angular scales greater than five arcminutes) were not included in the self calibration gain determination, preventing corruption from poorly modelled extended sources in the image.
    
    Many fields close to the Galactic Plane were difficult to image because of nearby bright and extended emission; the short RACS observations have limited capacity to represent these well (see Section \ref{subsec:uvsampling}). Of the 146 fields within ten degrees of the Galactic Plane, 67 were imaged without visibility data from baselines shorter than 35\,metres.  Solar emission also caused similar difficulties for some fields: 63 fields were imaged with baselines shorter than 100\,metres excluded. The minimum baseline length used for each RACS image is recorded and released in the survey database (see Section \ref{sec:datarelease}).

    \item {\bf Calibration application : } gain adjustments from the self-calibration and from polarisation leakage calibration were applied to the visibility data, which can then used for any further imaging.
    \item {\bf Mosaicing : } all beam images were combined into a single image for the whole tile; at each point the final image is a linear combination of the overlapping beam images using the primary beam model described below.
\end{itemize}


\subsection{Flux-density calibration}
\label{subsec:fluxscalecorr}

The ASKAP intensity scale is tied to the flux-density of PKS\,B1934$-$638 whose spectral energy distribution is described by \cite{Reynolds:1994vd}; the scale is set independently for each beam with an observation of that source.  ASKAP images are mosaics of the 36 beam images.  Over the course of the RACS observations we have had incomplete knowledge of the beam reception patterns, and the efficacy of the flux-density calibration.  Although the mosaicing operation assumes regular and identical beam shapes, we know that to be an approximation.  The disposition of calibration and science observations varies with respect to their relative times and elevations in the sky; the precise effect of these variations on the flux-density calibration is unknown.

In the light of these, we seek to quantify the flux-density scale as functions of position $(l,m)$ in the footprint and of observation time $t$. We compare total flux-densities $S_{\text{RACS}}$ of sources in each RACS image with that of their counterparts in either or both of the SUMSS and NVSS catalogues.  We form ratios $R_x = S_{\text{RACS}}/S_x$ where $x \in [\text{SUMSS},\text{NVSS}]$ and suppose that 
\begin{align}
    \label{eqn:fluxcal1}
    R_x &= \left( \frac{f_{\text{RACS}}}{f_x} \right)^{\alpha_x} F(l,m,t)
\end{align}
where $f$ is the radio-frequency of the survey, $\alpha_x$ is the characteristic source spectral index between the two survey frequencies and $F$ describes the unknown variations in flux-density scale.  We make the simplifying assumption that 
\begin{align}
    \label{eqn:fluxcal2}
    F(l,m,t) &\simeq F_1(l,m) F_2(t)
\end{align}
is the product of two independent functions.  Cast in this way, estimations of functions $F_1$ and $F_2$ can be made without knowing the values of $\alpha_x$, provided we retain reliance on the absolute scale determined from PKS\,B1934$-$638. Specifically, we assume that averaged over time, the flux-density scale derived from the calibration data is accurate and free from systematic error.

Until recently, measurements of the beam pattern have been insufficiently repeatable to assume anything but a simple average shape for all beams: a circular gaussian with width at half-power of $\text{FWHM} = 1.09\lambda/D$, where $D$ is the antenna diameter and $\lambda$ is the wavelength of the observed radiation. All ASKAP images, including RACS images, have been produced by linearly mosaicing beam images using that assumed pattern.

Analysis of source flux-densities in RACS images has revealed inconsistencies: (i) in comparison to other standards, flux-densities measured in RACS images were $\sim$10 percent too high, and (ii) there were clear indications that this discrepancy was not uniform, but dependent on source position within a tile. We were led to review the standard mosaicing procedure and the transfer of flux-density scale from the primary calibrator, as described below.

\subsubsection{Flux-density comparisons}
Comparison of RACS source flux-densities with values recorded in the SUMSS and NVSS catalogues has followed a procedure designed to make an unbiased estimate of the RACS flux-scale anomaly. To avoid the complexity of accounting for the variable PSF (see Section \ref{subsec:psf}), a large subset of RACS images were convolved with a gaussian function to give each a uniform (and circular) PSF with width 25\,arcseconds. The image analysis tool \textsc{Selavy} \citep{2012PASA...29..371W}, which uses the Duchamp source-finding algorithms \citep{2012MNRAS.421.3242W}, was used (with its default parameters) to identify sources in these images.  Comparison sources were chosen to have a nearest-neighbour separation $\Delta\theta_{N} > 90$\,arcseconds, a separation from SUMSS or NVSS counterpart of $\Delta\theta_{m} < 10$\,arcseconds, and a signal-to-noise ratio $\text{SNR} = S_p/\sigma \geq 10$, where $S_p$ was the peak flux-density and $\sigma$ the root-mean-square ({\em rms}) image noise at the source location. To ensure that chosen sources were unresolved, we used the following procedure (see also paper II). For all single component sources within a Selavy catalogue we examined the total to peak flux-density ratios $r_S = S_t/S_p$. For unresolved sources the distribution of this quantity should have unit mean  and variance proportional to $\text{SNR}^{-2}$.  Resolved sources appear as positive outliers. We determine the $5^{th}$ percentile as a function of $\text{SNR}$ to define a lower envelope to the distribution \citep{2008ApJ...681.1129B, 2017A&A...602A...1S}, which we describe as $r_{S_L}  = 1 - A \times \textrm{SNR}^{-B}$ where $A$ and $B$ give the best fit. This allows an estimate of the distribution's upper envelope, which is obscured by the presence of resolved sources, to be estimated as  $r_{S_U}  = 1 + A \times \textrm{SNR}^{-B}$.
All sources with $r_{S_L} < r_S < r_{S_U}$ were defined as unresolved, which led to a total of 67,595 sources identified for the following analysis.

\subsubsection{Position-dependent variations}
\label{subsubsec:pos_var}
\begin{figure*}
	\includegraphics[width=\linewidth]{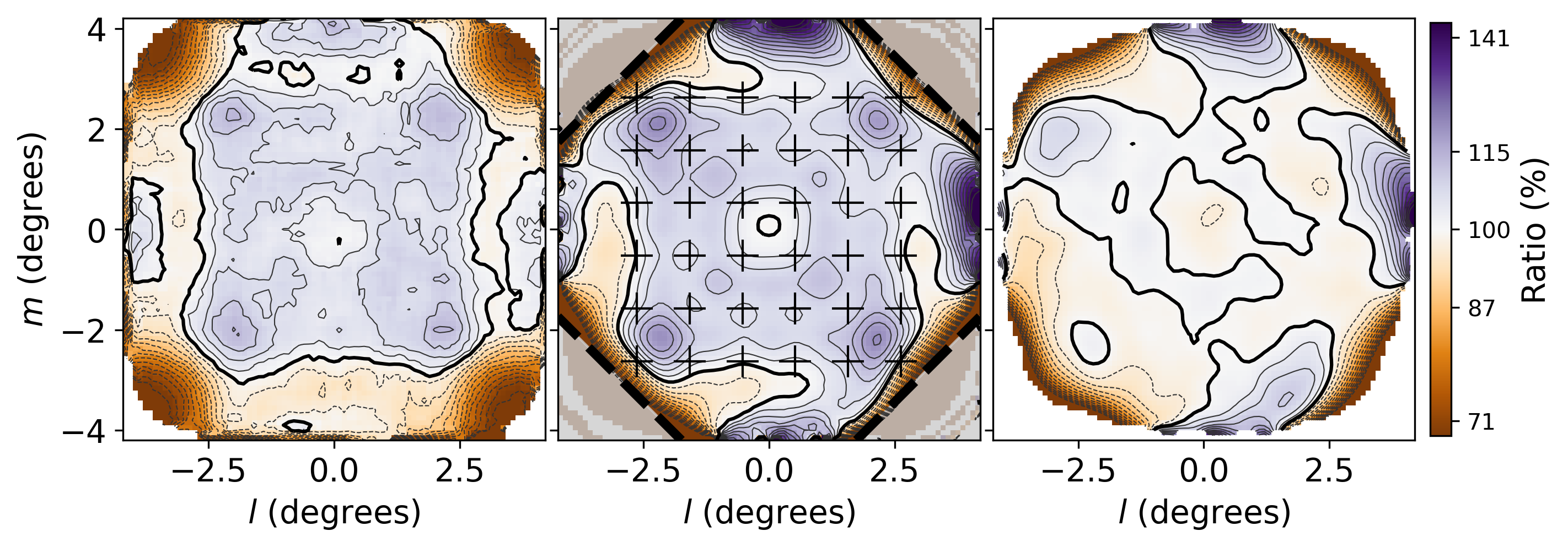}
	\caption{Brightness calibration over the field-of-view. Left: The flux-density ratio of RACS sources (uncorrected) to their counterparts in the SUMSS and NVSS catalogues, determined as described in the text (Section \ref{subsec:fluxscalecorr}). Centre: The direction-dependent correction $C_0$ derived from holography beam measurements and equation \ref{eqn:fluxcorr}; the corners of the field, outside the dashed lines, were not sampled by the holography procedure; beam positions are marked with a $+$. Right: The extra multiplicative factor $C_1$ used with the values in the centre panel to correct the flux-density scale variations; it shows the extent to which the holography-based correction matched the flux-density comparison data---very well across most of the field, and poorly near the edge of the holography grid.
	In all panels the contour interval is constant in the logarithm such that adjacent contours differ by a factor of 1.03, with the heavy contour at 1.0.} 
    \label{fig:flux_correction}
\end{figure*}

To quantify the direction-dependent variation of flux-density scale, comparison ratios ($S_{\text{RACS}}/S_{\text{SUMSS}}$ and $S_{\text{RACS}}/S_{\text{NVSS}}$) of the source sample were binned according to the source positions $(l,m)$ relative to the optical axis, and the median determined for each bin. Only the relative fluctuations over the field of view were needed, so both sets of ratios were scaled to have mean 1.0.  The result is an estimate of $F_1(l,m)$ in equation \ref{eqn:fluxcal2} and is shown in the left panel of Figure \ref{fig:flux_correction}. It shows a four-fold symmetry across the field, suggestive of incorrect primary beam correction.



\begin{figure}
	\includegraphics[width=\columnwidth]{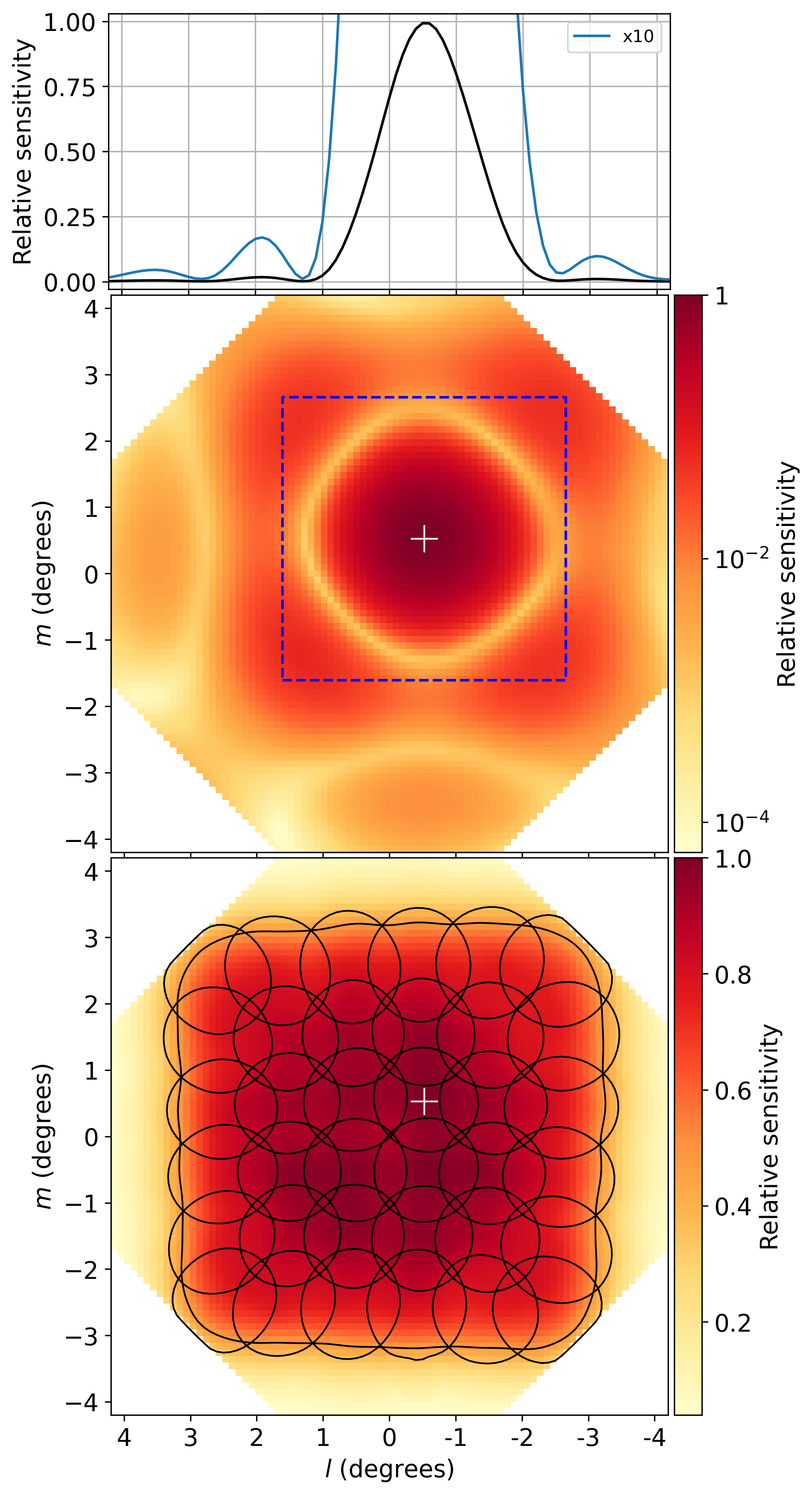}
	\caption{Primary beam sensitivity patterns determined by holography. The central panel shows, on a logarithmic scale, the sensitivity pattern for beam-0 at 930\,MHz, which in the RACS footprint is offset 0.525\degr from the field centre in both cardinal directions. This display clearly shows the corner regions of the field not sampled during the holography measurements. The dashed square marks the boundary of the image made for beam-0. The top panel shows the variation of sensitivity on a linear scale, horizontally through the beam centre. The bottom panel shows all beams as contours at their half-power level.  An estimate of the sensitivity of the 36-beam mosaic is indicated by the density of shading, with the outer contour drawn at the 50 percent level. The white cross in bottom and central panels marks the intended centre of beam-0.} 
    \label{fig:holo_beams}
\end{figure}

\paragraph{Beam shape measurements}
\label{para:holography}
Beam patterns for the RACS footprint were measured on 2020FEB06 using holography and the procedure described by \cite{Hotan:2016uk}. The new measurements were significantly more consistent than earlier attempts, showing less antenna to antenna variation in beam shape. We attribute this improvement to the advances made over the past year in phased-array setup and calibration. 
The measurements yield complex voltage measurements as a function of frequency for each beam on each antenna (apart from a reference antenna) on a grid of points covering the field-of-view.  For our application, a robust array-wide mean power pattern was constructed for each beam, resulting in 36 beam maps for each of the 288 frequency channels.  Interpolation was used to span gaps in the spectrum caused by radio-frequency interference. Figure \ref{fig:holo_beams} shows the mean pattern for beam-0 over 26 of the 32 measured antennas (the missing four were the reference antenna and three others that were unavailable during the holography observations) and the measured shape at half-power of all beams in the footprint. All of the 26 antennas used for the array-wide mean had power patterns differing from the mean by less than 4 per cent.  The antennas not used in the mean pattern were the outer six; the holography procedure uses the source Virgo A as a reference and because of its angular size, it has insufficient power on the longest ASKAP baselines to make reliable beam measurements on those outer antennas. 

\paragraph{Flux-density scale correction}
\label{para:fluxscalecorrection}
If $B_m$ is the image brightness determined using the standard reception pattern assumption $A_i$ for each beam, and $H_i$ is the pattern as measured by holography, then
\begin{align}
    \label{eqn:fluxcorr}
    B_t &= B_m \frac{\sum A_i^2}{\sum c_i  H_i A_i}
\end{align}
is the true brightness. In equation \ref{eqn:fluxcorr}, $c_i$ corrects for the effect of a coma-related asymmetry in beam-shape moving the peak response away from each beam's nominal position.  Ideally, without coma, $c_i = 1.0$. However, we measured values of $0.92 < c_i < 1.0$, the most extreme values for outer beams at the low frequency end of the band. This was a major cause of the high flux-densities measured in RACS images.  Appendix \ref{app:tt1corr} describes the derivation of equation \ref{eqn:fluxcorr} and its use to correct both the zeroth and first Taylor term images.

Figure \ref{fig:flux_correction} (centre) shows the evaluation of the factor $C_0 = B_t/B_m$ in equation \ref{eqn:fluxcorr}. No holography measurements were made outside the heavy dashed lines in that Figure.

Note that this flux-density comparison was constructed from observations made over many months, most of them several months before the holography measurements.  The degree of similarity between the left and centre panels indicates a pleasing level of stability of ASKAP beam reception patterns.  In future, holography beam measurements will be made routinely as part of the instrumental flux calibration, and the values of $H$ will be used directly by the linear mosaicing software so avoiding the use of equation \ref{eqn:fluxcorr} and the procedure described here.

Although there is clear similarity between the forms of the measured and predicted intensity scale errors, the calculated factor $B_t/B_m$ is insufficient for a full correction of the RACS image intensity scale.  Both the incomplete holography measurements and the probable failure of the single measurement to represent actual beam reception patterns earlier in the survey contribute to this inadequacy.  Therefore, we have used the same set of flux-density ratios of the 67,795-source sample described above to determine a multiplicative adjustment $C_1$ to improve the correction is areas of the field poorly characterised by holography. Flux-density ratios were again binned according to position $(l,m)$; median values were determined for each bin and normalised to give a mean value of 1.0 over the area of valid holography measurements. The normalisation was done independently for the $S_{\text{RACS}}/S_{\text{SUMSS}}$ and $S_{\text{RACS}}/S_{\text{NVSS}}$ ratios, allowing the two surveys to be used in this way in spite of their different centre frequencies. By forming the flux-density correction in this way---constraining the mean correction over the holography field and using the source sample drawn from most of the RACS survey area $-84$\degr$<\delta <+30$\degr, (SUMSS: $\delta < -30$\degr; NVSS: $-40$\degr $< \delta < +30$\degr), we have retained the link to the original flux scale calibration with  PKS\,B1934$-$638.  Thus, any change in the mean flux-density measurements is a result of the improved primary beam models, not the comparison with SUMSS and NVSS. The factor $C_1(l,m)$ is shown in the right-hand panel of Figure \ref{fig:flux_correction}. The correction applied to all RACS images was $F_1(l,m) = C_0 C_1$.

\subsubsection{Time-dependent variations}
\begin{figure*}
	\includegraphics[width=\linewidth]{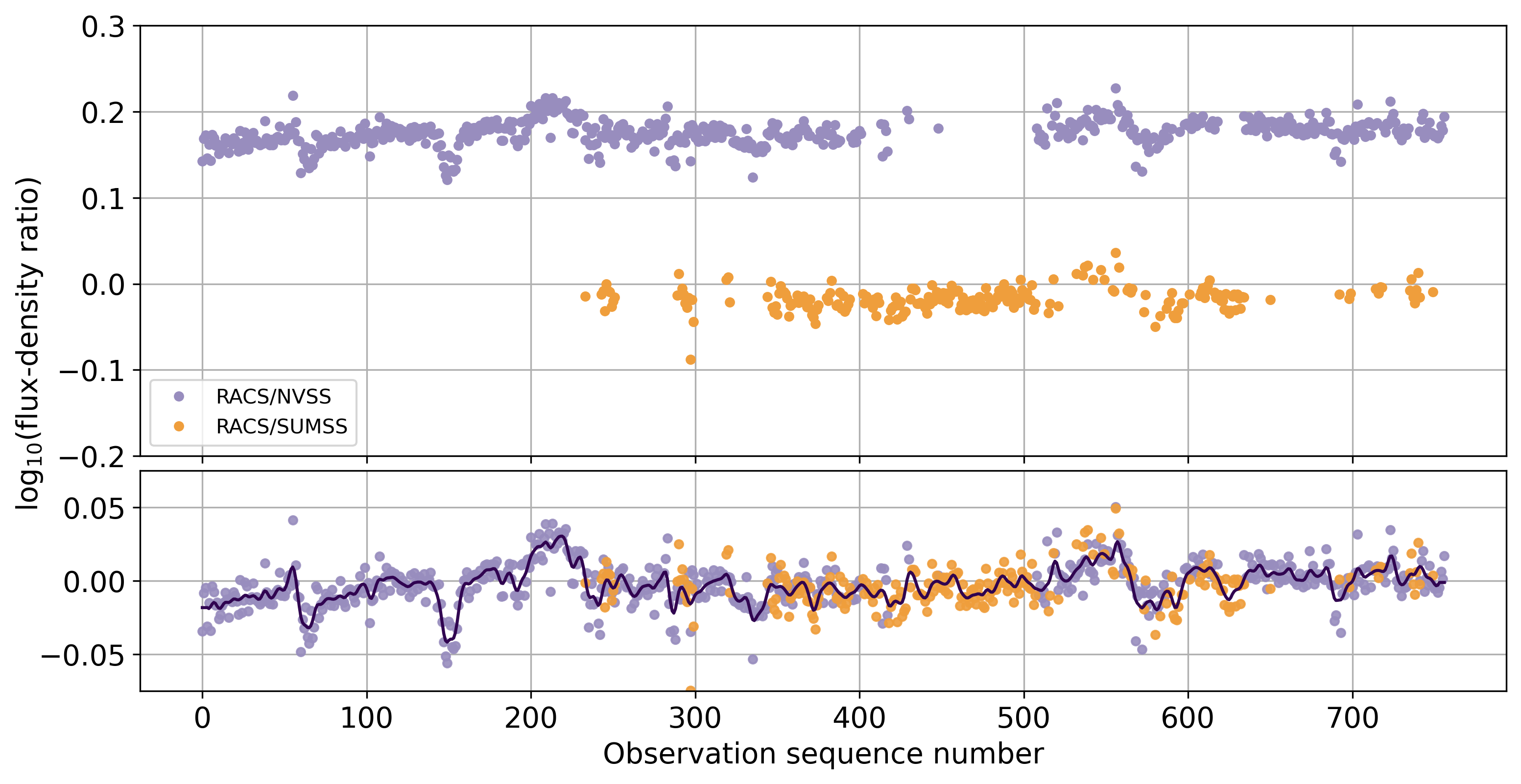}
	\caption{Apparent flux-scale variations. The abscissa in both panels is the sequence number of each observation. The upper panel shows, for each tile, the median flux-density ratio between RACS and NVSS and SUMSS measures as their logarithms. No allowance has been made for the different frequencies of the surveys; typical source spectral indices shift the two sets of points away from unity---$S_{\text{RACS}}/S_{\text{SUMSS}} < 1$ and $S_{\text{RACS}}/S_{\text{NVSS}} > 1$. The lower panel has both sets of ratio logarithms shown with their means subtracted.  The ratio variations appear to have a random component added to a slowly varying component with typical scale of 10--20 observations, and shown as a dark line.} 
    \label{fig:flux-obs-sequence}
\end{figure*}

With the large offset-dependent scale errors removed, we examined the data for evidence of other forms of systematic error.  Such errors may be expected from a number of factors such as a dependence on the zenith angle of the observation, temporal and angular distance from the calibration observation and variation in receiver gains.  Figure \ref{fig:flux-obs-sequence} shows the variations in apparent intensity scale against the observations' position in their time-ordered sequence. The separation between RACS/SUMSS and RACS/NVSS points in the top panel arises from the different survey frequencies and the typical negative spectral slope of radio sources. As with the position-dependent analysis above, we are interested in changes in the RACS flux-density scale, not in its mean value and so the bottom panel shows the two series with the mean removed and is an estimate of $F_2(t)$, equation \ref{eqn:fluxcal2}.  There is evidence of a non-random component in the variation: note the similarity in fluctuations between the two series in the observation sequence range [500,600].

As a trial, we have computed an extra scale factor for each observation from a smoothed form of this series. When this was applied to the images there was no clear improvement: no decrease in the overall variance of flux-scale and no significant change in the implied average spectral index between RACS sources and counterparts in catalogues formed at other frequencies.  Therefore, and because we do not yet understand the physical origin of the non-random scale variation, we have not made any further correction to RACS image intensity scales.

\subsection{Image quality}
\label{subsec:imagequality}
In this section we present quality assessments for the first epoch RACS observations: image noise and sensitivity, astrometric precision and accuracy and photometry---the flux-density scale. We also give some general comments on the image fidelity, the degree to which images are representative of the true sky brightness.  We leave thorough analyses of survey completeness and reliability to the RACS paper II currently in preparation. 

\subsubsection{Image noise}
\label{subsubsec:noise}
\begin{figure*}
	\includegraphics[width=\textwidth]{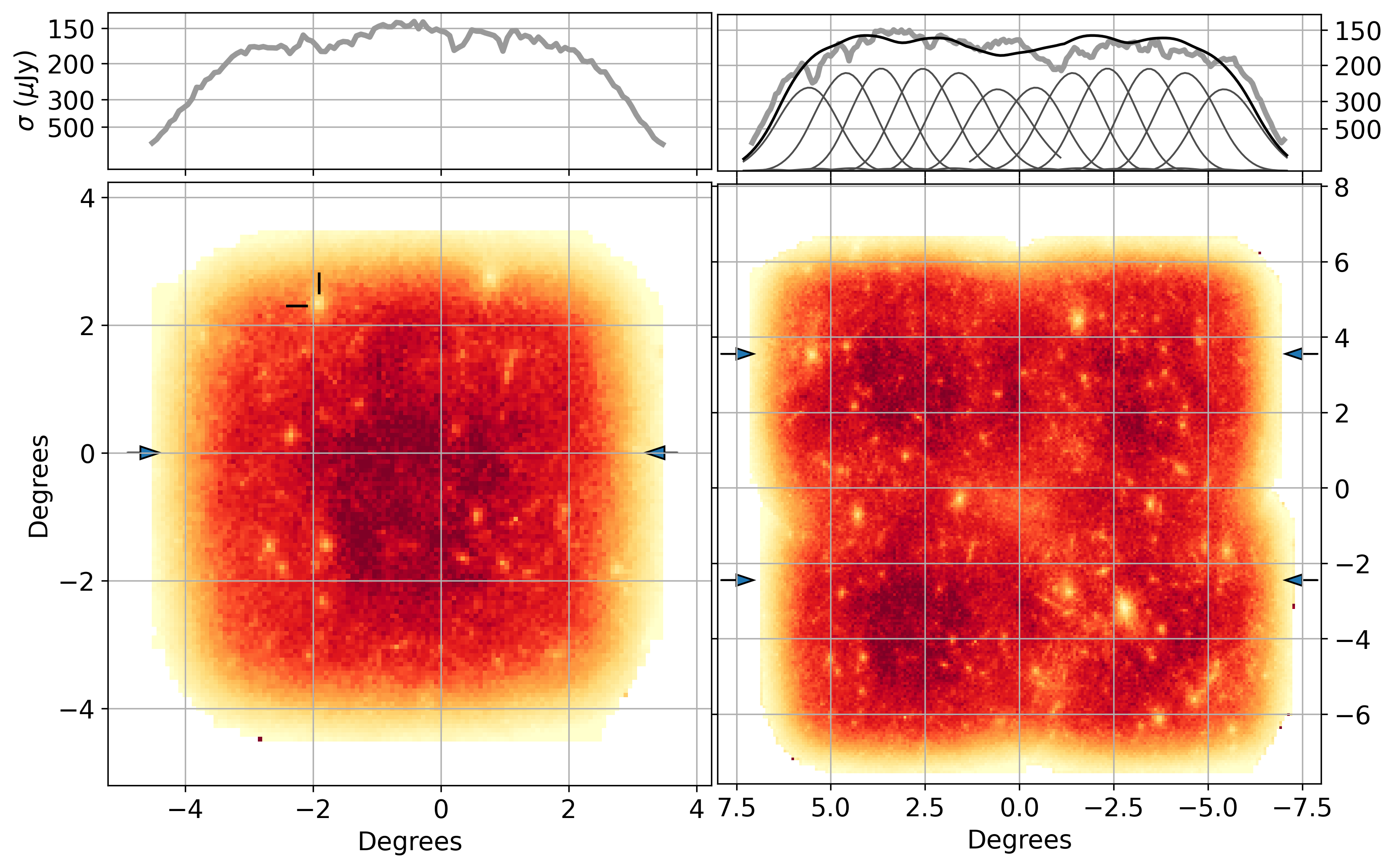}
    \caption{Image noise, displayed as the inverse {\em rms} ($1/\sigma$) computed as described in the text (Section \ref{subsubsec:noise}). For each, the upper panel shows a horizontal profile (grey line) averaged over the range indicated by the arrows beside the images. Left: The variation of noise amplitude across a single Stokes-I image tile; the central value is $\sigma \simeq 150 \mu$Jy. This tile is colocated with the bottom left tile of the right-hand panel. The feature marked is associated with the source PKS J1102$-$0951 as described in the text.  Right: The variation of image {\em rms} over a mosaic of four tiles. These tiles are centred at 1033-06, 1057-06, 1031-12, 1056-12. In the top right panel we also show the beam profiles along a horizontal path through beam-0 (between the arrows). The beam profiles are taken from the holography measurements described in Section \ref{subsubsec:pos_var}, and weighted by a system noise estimated for each beam from the flux-density calibration data record of PKS\,B1934--638. The beams shown are numbered 23, 8, 1, 0, 15, 34. The smooth black curve is the resultant sensitivity expected from these beams.}
    \label{fig:tile_rms}
\end{figure*}

For each image we generated a noise map by dividing the image into cells of $100\times100$ pixels and in each, measuring the signal spread about the mode, allowing an estimation of the {\em rms} noise without contamination by compact sources in the image. Figure \ref{fig:tile_rms} shows some example noise images.  These views of the survey products show clearly the variation of sensitivity across the field-of-view. They also show that for these observations there is little sensitivity ripple imposed by the grid of beams within the tile. The upper right panel of Figure \ref{fig:tile_rms} indicates the beam positions and width (at the centre frequency). The right panel also shows the variation of sensitivity across the tile boundaries.

Figure \ref{fig:tile_rms} also illustrates another feature of the images: increased noise close to bright sources.  For example, the noise peak seen in the upper left of the left panel surrounds PKS J1102$-$0951, a source 
with a flux-density of 0.9\,Jy in the RACS image.

\begin{figure}
	\includegraphics[width=\columnwidth]{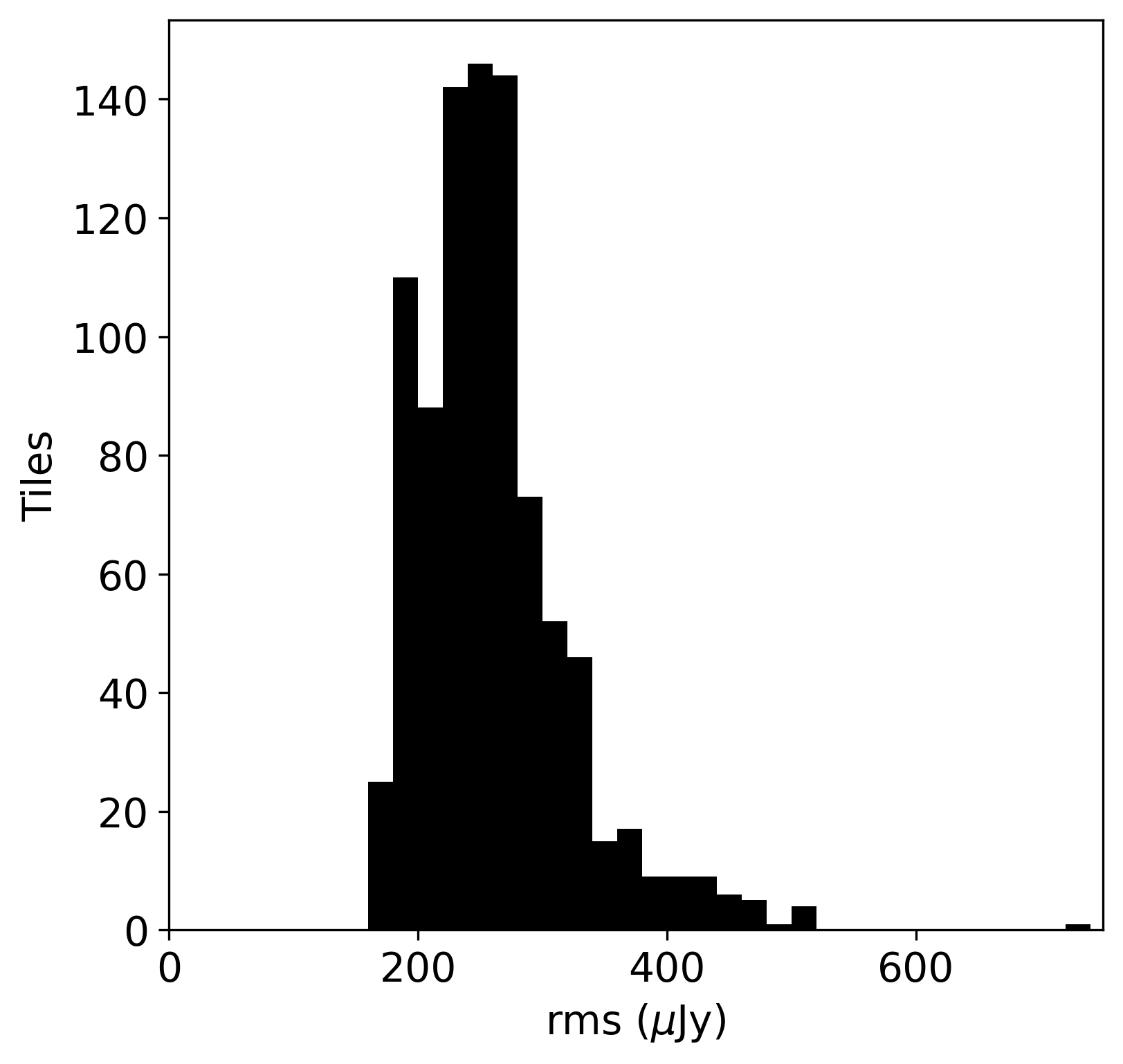}
    \caption{The distribution of the median {\em rms} values for the 903 tile images in the first RACS image release.}
    \label{fig:rms_hist}
\end{figure}

\begin{figure*}
	\includegraphics[width=\linewidth]{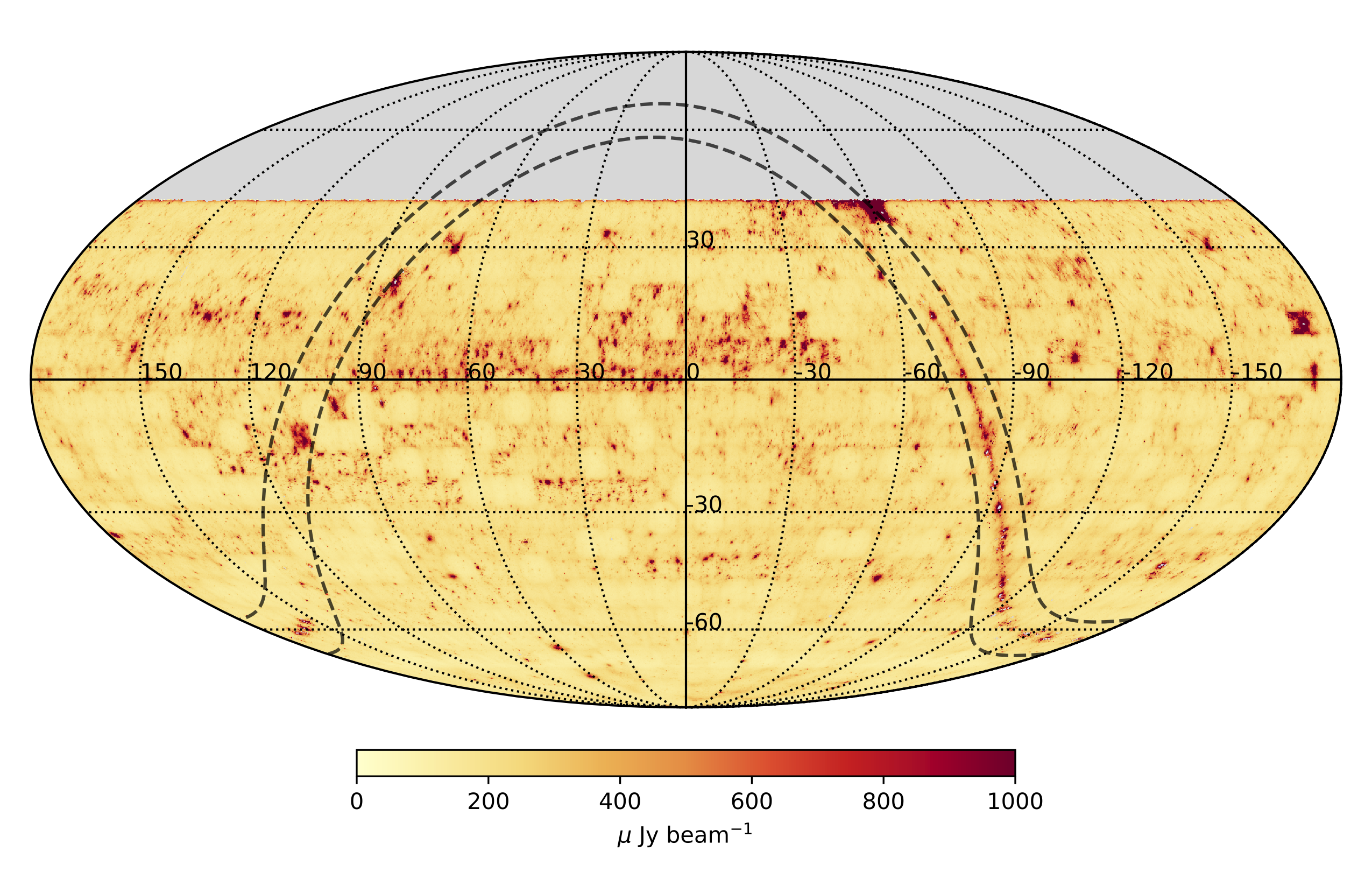}
	\caption{Image noise across RACS survey area, derived from the 903 {\em rms} images described in Section \ref{subsubsec:noise}. The Galactic plane is visible over part of its longitude range. The prominent area with high noise near right-ascension -170\degr\ and declination =12\degr\ is the tile containing the source Virgo-A, which is bright and has complex structure.}
    \label{fig:all_tiles_rms}
\end{figure*}

Figure \ref{fig:rms_hist} shows the distribution of rms noise values for RACS Stokes I images. The median {\em rms} noise is $\sigma_\text{med} = 250\,\mu$Jy/beam and $90\%$ of RACS images have $\sigma_\text{med} < 330\,\mu$Jy/beam. The median rms of all RACS images is shown in Figure~\ref{fig:all_tiles_rms}.

\subsubsection{Point-spread-function}
\label{subsubsec:psf}

As mentioned in Section \ref{subsec:psf}, the variation in PSF size across a single tile mosaic can be significant because of large baseline foreshortening gradients across the field of view, and also  beam-specific data losses due to instrumental problems. The latter became less acute through the observing period because of improvements in the telescope reliability. Although the differential foreshortening effect can be minimised by observing close to the meridian, some variation of PSF size and shape will always be a feature of ASKAP images.

We draw attention to the impact PSF size variation on photometric interpretations of the image. The intensity in an astronomical image is expressed in units of flux-density per unit solid-angle; in the case of radio images the traditional measure is mJy/beam where ``beam''  stands for the solid-angle $\Omega$ subtended by the main lobe of the PSF. In RACS images, $\Omega = \Omega(l,m)$ varies across the image so we must declare a reference PSF with solid-angle $\Omega_{\text{ref}}$. In practice this reference is the PSF of the beam-0 image, and its parameters are recorded in the mosaiced image header. Then the total flux-density $S$ of a source at position $(l,m)$ in the image $I$ is
\begin{align}
\label{eqn:flux_integral}
S = \frac{\Omega_{\text{ref}}}{\Omega(l,m)} \int_{\text{src}}I d\Omega
\end{align}
The expression is only applicable for computing the total flux-density of a source; the brightness scale, including the peak flux-density of sources is correct without adjustment.

To support use of the RACS image set, we release two forms of each image. First we provide the mosaic produced by the pipeline as described  in Section~\ref{subsec:calibration-and-imaging}.  These images provide the highest, but variable, spatial resolution. With knowledge of the PSF variations over these images, source flux-densities can be determined using equation \ref{eqn:flux_integral}. Each beam image is restored using an analytic (elliptical gaussian) approximation of the main lobe of the PSF, and then used to form the 36-beam mosaic. Therefore at each point $(l,m)$ in the mosiac, the effective or resultant PSF is the linear combination of the beam-specific gaussian functions, and we can approximate it with another elliptical gaussian whose volume is the PSF area $\Omega(l,m)$ in equation \ref{eqn:flux_integral} above.  With each variable-resolution RACS image we publish its PSF area as a function of position in a separate file, enabling correct total flux densities to be determined by source finding software that uses the reference PSF information carried in the image file.  Note that in some areas (for example around the celestial pole) the PSF of adjacent beams can differ markedly, making the effective PSF in the mosaic (the weighted mean of PSFs of overlapping beams) non-Gaussian and the results of source-shape fitting less reliable.

The second image type released is formed by mosaicing  beam images that have been convolved to a common spatial resolution\footnote{Using \url{https://github.com/alecthomson/RACS-tools}}. These images may lose some  spatial resolution, but can be used to estimate source flux-densities without the need for equation \ref{eqn:flux_integral}.   For each RACS tile, using the methods provided by \texttt{radio-beam}\footnote{\url{https://github.com/radio-astro-tools/radio-beam}}$^{,}$, we first find the smallest common PSF i.e. the PSF with the smallest area to which the set of 36 varying PSFs can be convolved. Second, for each beam $i$ we compute both the convolution kernel $k_i$ required to transform $\text{PSF}_i$ to that common resolution, and the scaling factor $f_i$ required to maintain the image units of mJy/beam. We then produce common-resolution beam image $I_{\text{cmn},i }$ as the convolution of variable-PSF image $I_{\text{var}}$ with the convolving beam kernel:
\begin{align}
\label{eqn:cmn-res}
I_{\text{cmn},i }= f_i \left(  k_i * I_{\text{var}, i} \right),
\end{align}
for all beams in a tile ($i = 0,...,35$). These common-resolution beams are then mosaiced as described above.

\subsubsection{Astrometry}
\label{subsubsec:astrometry}
The astrometric performance of ASKAP (and all synthesis radiotelescopes) relies on the phase tracking stability of the receivers and on the quality of phase reference measurements.  ASKAP maintains a phase-centre for each beam and so has 36 phase tracking systems tied to the observatory frequency standard \citep[a rubidium atomic clock; ][]{Hotan:2020}.  RACS observations are referred to the phase observed during an observation of PKS\,B1934--638. In this section we write astrometric errors as $\boldsymbol{\epsilon} = (\Delta\alpha\cos{\delta}, \Delta\delta)$, a vector with components in the Right Ascension and Declination directions.

\begin{figure}
	\includegraphics[width=\linewidth]{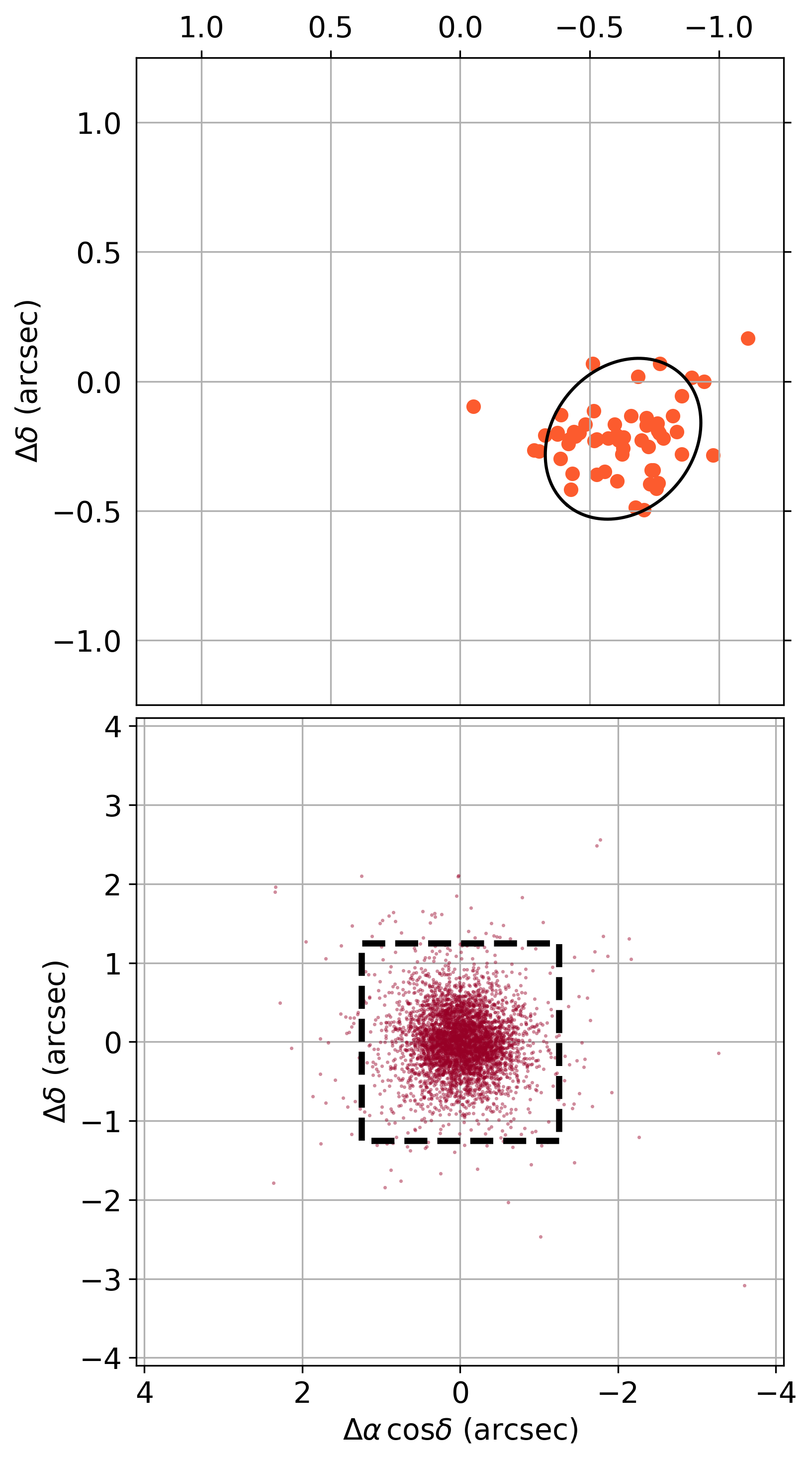}
	\caption{Beam-to-beam offsets. Upper: The apparent position differences of 57 sources observed in both beams 2 and 14 for the field centred on 00$^h$00$^m$\,$-$37\degr38$'$. The sources were selected to be compact and exceed a signal-to-noise ratio of 20, their spread consistent with expected error ellipse drawn in black. The upper panel is the size of each image pixel (2.5 $\times$ 2.5 arcseconds). Lower: The distribution of mean beam-to-beam offsets over 16 beam pairs in 567 RACS fields.  In most cases the beam-to-beam astrometric difference is less than the width of a 2.5 arcsecond pixel (dashed box).}
    \label{fig:astrometric_precision}
\end{figure}

\paragraph{Astrometric precision}  The overlap of ASKAP beams on the sky provides a means for assessing astrometric precision as there are typically fifty suitable sources that are common to any pair of adjacent beams. For each of these sources we have two position measurements and the distribution of their differences is observed to have both a random and a systematic component. Figure \ref{fig:astrometric_precision} (top) shows the position difference distribution for a single pair of adjacent beams in the field centred on 00$^h$00$^m$\,$-$37\degr38$'$. The scatter in position is consistent with the expected position variance of
\[
\sigma(\epsilon)^2 \simeq \Delta\theta^2 \times 2 / (4 \ln 2 \times \text{SNR}^2)
\]
\citep{1997PASP..109..166C}; $\Delta\theta$ is the angular size of the PSF, $\text{SNR}$
is  the signal-to-noise ratio of the source and the factor of 2 arises from differencing samples from two independent distributions. In this case the PSF size is 16 $\times$ 13 arcseconds and comparison sources have $\text{SNR} > 20$. The expected error ellipse is drawn in the figure, centred on the mean position.

The difference distribution has a non-zero mean: there is a systematic shift in apparent position difference of sources depending on which pair of adjacent beans is considered. The lower panel of Figure \ref{fig:astrometric_precision} shows the distribution of systematic shifts over all independent beam-pairs for a sample of 567 RACS fields. Almost all shifts are less than the image pixel size of 2.5 arcseconds.  We attribute this systematic error to the effect of self-calibration, which is used for each beam image to improve the estimate of antenna gains.  The improvement in relative antenna gain phases is achieved at the expense of introducing some uncertainty in absolute phase, resulting in small astrometric shifts.

\begin{figure}
	\includegraphics[width=\columnwidth]{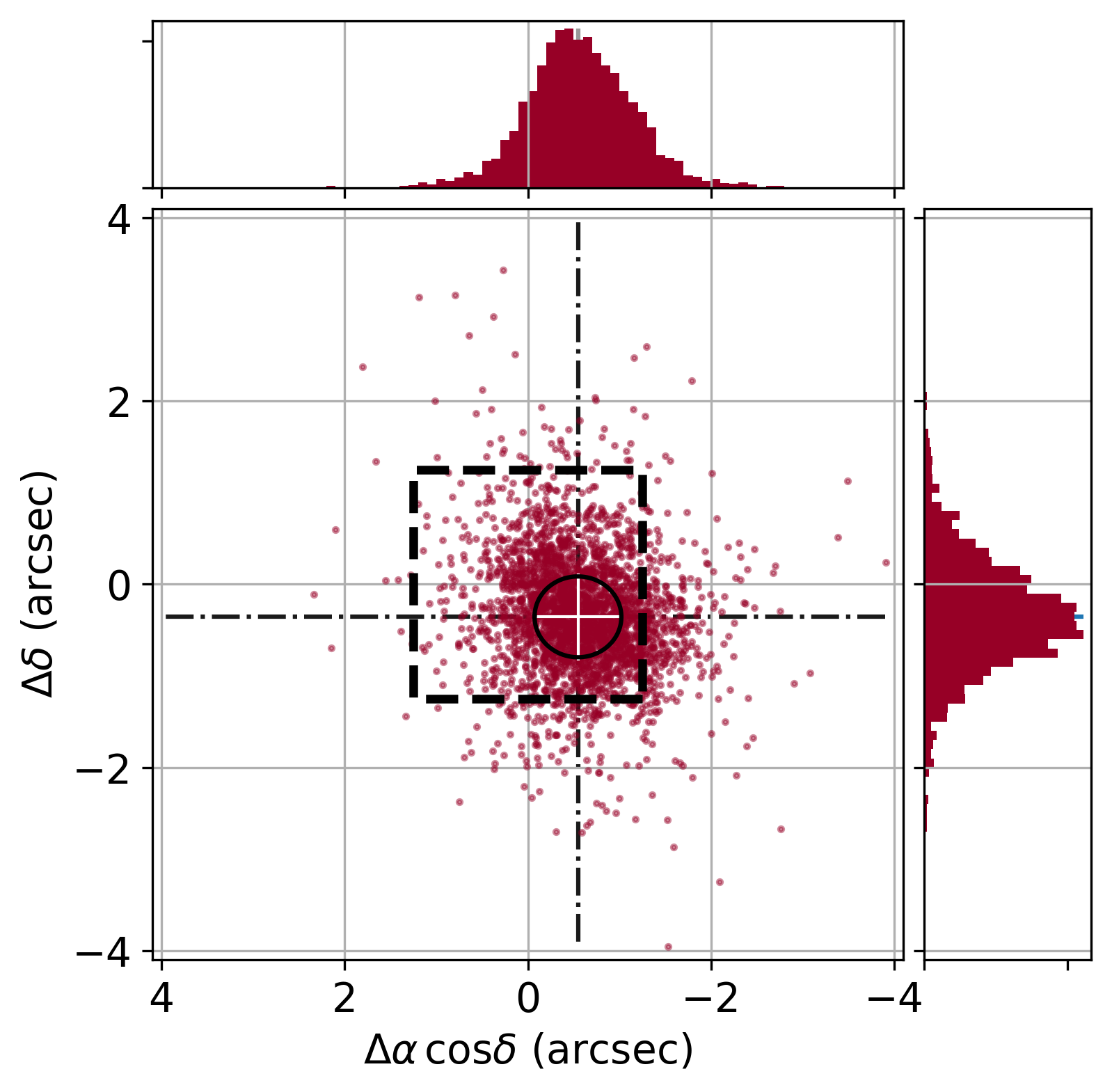}
	\caption{Position offsets of RACS sources relative to their ICRF counterpart. A systematic difference is observed; the median offset is $-$0.6 and $-$0.4 arcseconds in Right Ascension and Declination respectively.   The sources were selected to be compact and exceed a signal-to-noise ratio of 10. The text in Section \ref{subsubsec:astrometry} describes the basis for the size of the error ellipse drawn on the figure. The size of RACS image pixels is shown by the dashed square.}
    \label{fig:astrometric_accuracy}
\end{figure}

\paragraph{Astrometric accuracy} We have assessed RACS astrometry against the International Celestial Reference Frame (ICRF). The third realisation of the ICRF (ICRF3, Charlot et al. A\&A, {\em accepted}) is a catalogue of 4536 radio sources distributed across the sky; 3604 of these lie south of the northern limit of RACS.  At least one reference source appears in 506 RACS tiles. We matched ICRF and RACS sources, excluding matches for which the RACS source appeared resolved. The resulting 2915 position differences $\epsilon = s_R - s_{\text{ICRF}}$ are plotted in Figure \ref{fig:astrometric_accuracy}, where $s_R$ and $s_{\text{ICRF}}$ are the directions to the RACS and ICRF sources respectively.  The east-west and north-south components of $\epsilon$ are distributed with means and standard deviations $\Delta\alpha\cos{\delta} = -0.6\pm0.6$ and $\Delta\delta = -0.4\pm0.7$\,arcseconds. The error ellipse expected from Gaussian fitting considerations \citep{1997PASP..109..166C} is shown in the figure, centred on the distribution mean and sized according to the survey-wide mean PSF shape (14.9 $\times$ 13.7 arcseconds, position angle 53\degr) and $\text{SNR} \geq 10$, with an additional component determined empirically from Figure \ref{fig:astrometric_precision}. The cause of neither the mean position error nor its large variance is yet understood and will be the subject of future investigation.

\subsubsection{Photometry}
\label{subsubsec:fluxscale}
\begin{figure*}
	\includegraphics[width=\linewidth]{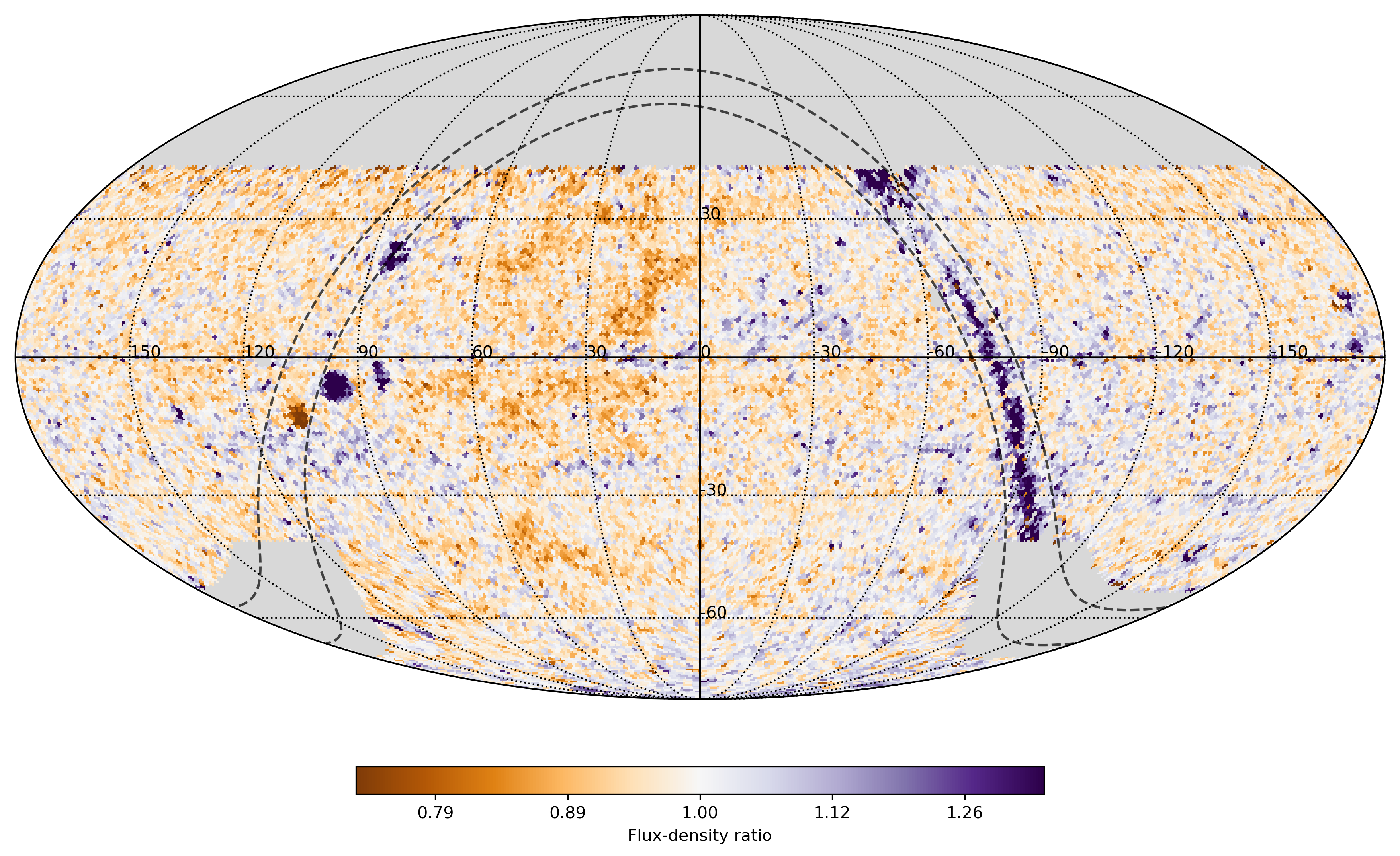}
	\caption{Apparent flux-scale variations over the sky, computed
	as described in the text. The Galactic Plane is visible as elevated values in the region covered by NVSS, and absent values over the SUMSS area. Galactic latitudes of $\pm$5\degr are shown.} 
    \label{fig:flux-variation-sky}
\end{figure*}

\begin{figure}
	\includegraphics[width=\columnwidth]{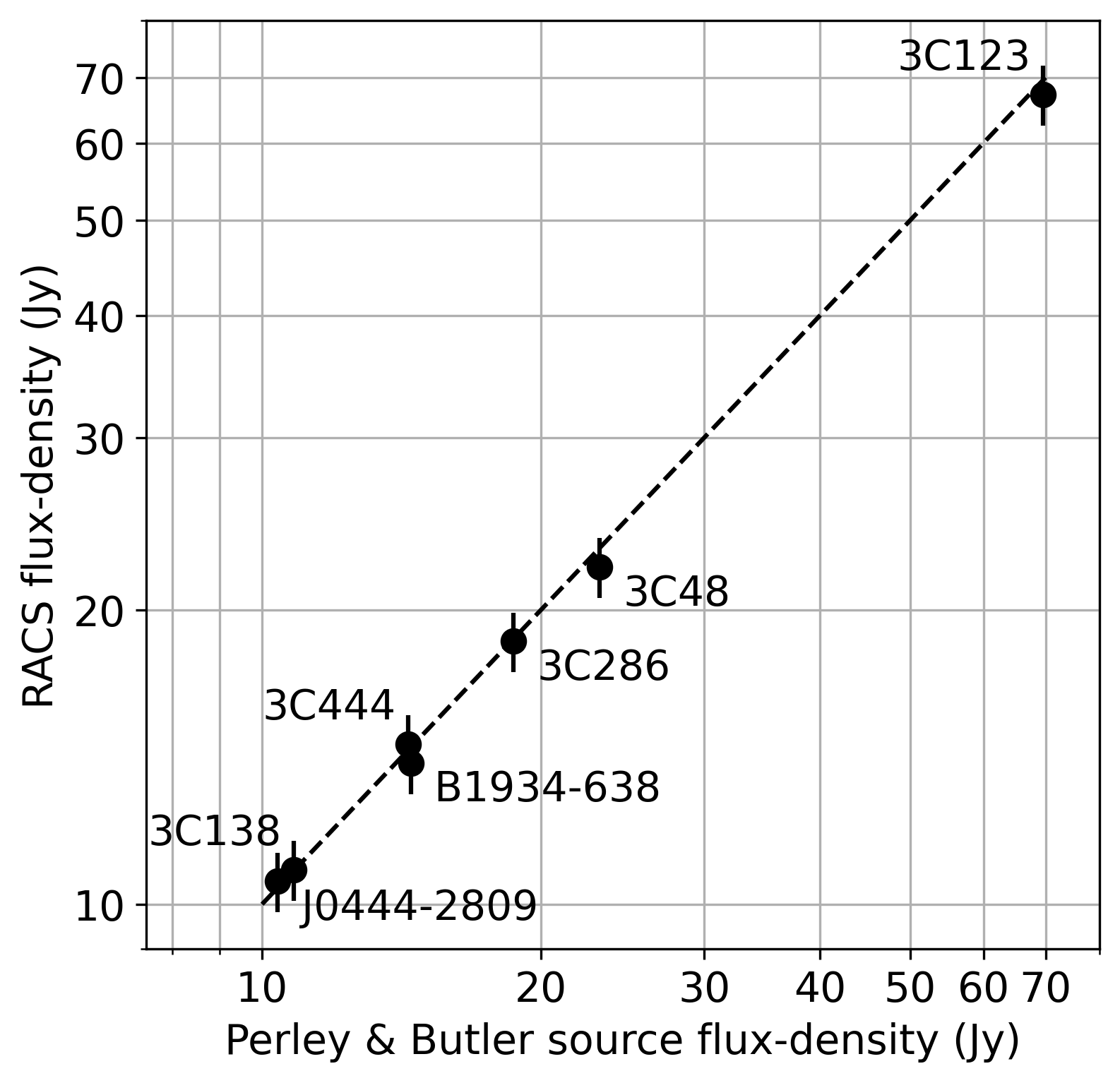}
	\caption{Comparisons of source flux-densities measured in RACS images to those from a set of well-characterised sources published by \cite{Perley:2017gn} and \cite{Reynolds:1994vd}. The error bars are computed from expression \ref{eqn:delta-s}.}
    \label{fig:flux_accuracy}
\end{figure}

In Section \ref{subsec:fluxscalecorr} above we discussed in detail the procedures that we devised to quantify and correct the intensity scale in RACS images. 
Figures \ref{fig:flux-variation-sky} and \ref{fig:flux_accuracy} summarise the photometry over the survey images being released.  The sky-wide comparison with SUMSS and NVSS catalogues is summarised in Figure \ref{fig:flux-variation-sky}, in which only the fluctuations about the mean flux-density ratios are shown, the same fluctuations shown in the top panel of Figure \ref{fig:flux-obs-sequence}.  The non-random component to the position-dependent variation, in part, is associated to the Galactic Plane (the SUMSS catalogue excludes sources within 10 degrees of the Plane). Other variations have unknown origin, but likely correspond to the variations discussed in Section \ref{subsec:fluxscalecorr} and illustrated in Figure \ref{fig:flux-obs-sequence}.

To quantify the uncertainty of source flux-density $\Delta S$ in RACS images we have analysed the values obtained for the set of sources that lie in the overlap between tiles and so have two independent measurements, each of which we assume to have the same uncertainty.  For each source $i$, let the two flux-density measurements be $^{a}S_i$ and $^{b}S_i$, with mean $S_i$. We analysed the distribution of their ratios $r_i = {}^{a}S_i/^{b}S_i$, determining the standard deviation of $r_i$ for a number of logarithmically spaced intervals in $S_i$. We found that the results were  well modelled by the expression $\Delta S_i = \Delta S_0 + fS_i$ in which $\Delta S_0$ and $f$ are constants across the survey: $\Delta S_0 = 0.5\text{\,mJy}$ and $f = 0.05$. We add to this a component of uncertainty to reflect the variations seen in Figure \ref{fig:flux-variation-sky}, which we believe is independent of the errors evident in the dual-measure analysis. Adding in quadrature the standard deviation of fluctuations displayed in Figure \ref{fig:flux-obs-sequence}, we arrive at a final expression 
\begin{align}
\label{eqn:delta-s}
\Delta S = 0.5\text{\,mJy} + 0.07S
\end{align}

To check the absolute flux-density scale we examined the RACS flux-densities of sources listed by \cite{Perley:2017gn}, which have well determined flux-densities over the RACS spectral range.  For each in the list, \cite{Perley:2017gn} indicate the extent of each source with a parameter LAS (Largest Angular Scale). We have chosen the six sources south of $\delta =$~+40\degr\ with $\text{LAS} \leq 120$\,arcseconds, and added PKS\,B1934$-$638, also with well determined spectral energy distribution \citep{Reynolds:1994vd}. Figure \ref{fig:flux_accuracy} shows the comparison; the error bar lengths were calculated from the expression \ref{eqn:delta-s} above. Several of the Perley\,\&\,Butler sources are well resolved in RACS, and their total flux-density was found as the sum of component flux-densities found by the \textsc{Selavy} image analysis tool.  Note that the likelihood of all seven sources falling within one standard-error of the expected value is very low ($\sim 0.68^{7} \simeq 0.07$), perhaps indicating that the value of $f$ in expression \ref{eqn:delta-s} is too high.

\subsubsection{Image fidelity}
\begin{figure*}
	\includegraphics[width=\linewidth]{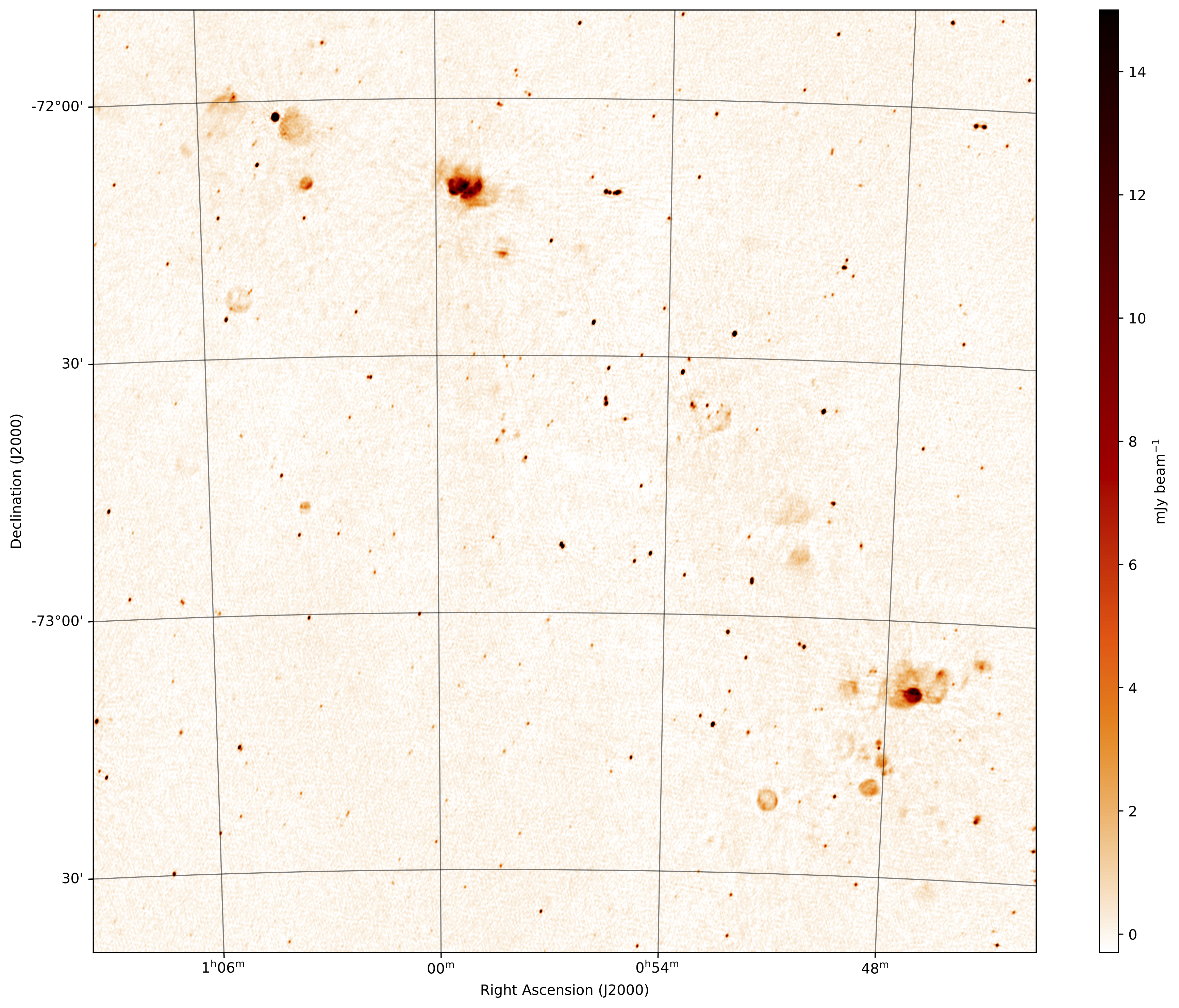}
    \caption{The RACS image of the Small Magellanic Cloud, as an example of a field with extended emission, with sources up to five arcminutes in size being well represented. The {\em rms} brightness is $\sigma \sim$~200$\mu$Jy/beam. The PSF has dimensions 18.5~ $\times$~11.5 arcseconds. The image has a dynamic range of about 3700:1. The bright nebula complex toward the top of the image is NGC 346 (also known as N66 and DEM 103), and to the lower right is N19 (DEM 32). A number of supernova remnant shells are visible, see, e.g.for example, \cite{2019A&A...631A.127M} for details.}
    \label{fig:smc}
\end{figure*}

\begin{figure*}
    \centering
    \includegraphics[width=\linewidth]{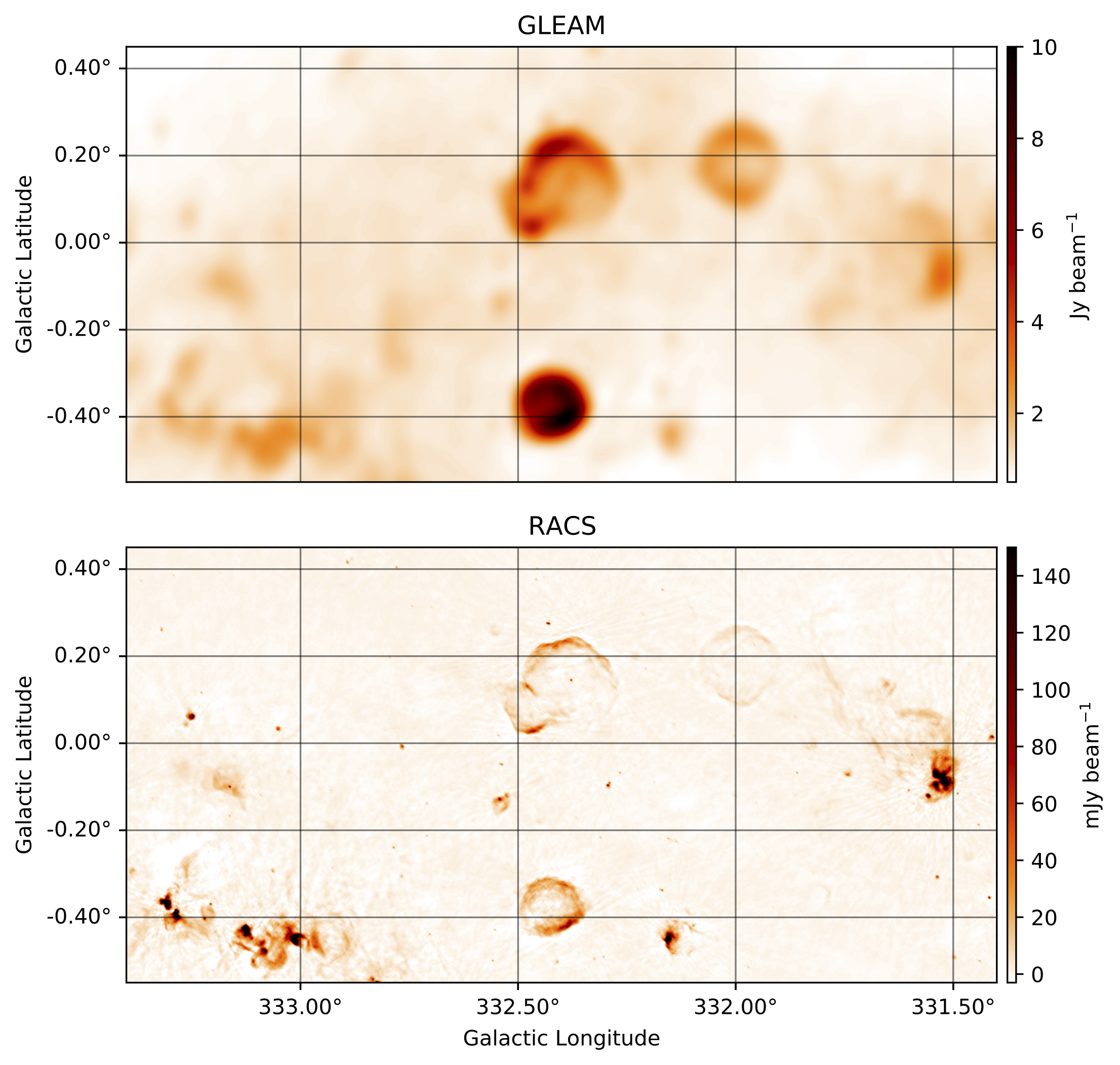}
    \caption{Images from GLEAM and RACS centred on Galactic coordinates $l=$332.3\degr, $b=0$\degr. This comparison illustrates the superior sensitivity of GLEAM images to large angular scales, and the finer resolution of RACS.  This RACS image has a PSF size 14.5$\times$11.5\,arcseconds and median image noise of 270\,$\mu$Jy/beam.
    The supernova remnants G332+0.2, G332.4$-$0.4 (RCW\,103), and G332.4+0.1 (MSH\,16$-$51, Kes\,32) are prominent \citep[cf][]{MSC}.} 
    \label{fig:racs_gp}
\end{figure*}

Radio images made with interferometers are typically able to well represent spatial scales from the size of their PSF up to a ``largest angular size'' $\Theta_{LAS}$ determined by the length of the shortest baseline $B_{min}$ as 
\[
\Theta_{LAS} = \lambda/(a B_{min})
\]
where $\lambda$ is the observing wavelength and $a$ is a number $1 \lesssim a \lesssim 2$, depending on the observation length and amount of Earth-rotation synthesis achieved. We make the distinction here between fidelity---the ability of the telescope to represent structures of a certain angular scale---and the telescope's sensitivity to those structures. Aperture-synthesis telescopes necessarily have less sensitivity to larger scales because of the decreasing collecting area on short baselines. However, structures smaller than $\Theta_{LAS}$ can be imaged provided they are sufficiently bright. Source components with size greater than $\Theta_{LAS}$ are not sampled by the telescope and so cannot be imaged. For the brief RACS observations made at $\lambda \sim 34\text{\,cm}$ we expect $25' < \Theta_{LAS} < 50'$. In regions close to the Galactic Plane we had difficulty properly calibrating the shortest baselines and used only those longer than 35\,m, so that in these images we expect $15' < \Theta_{LAS} < 30'$.  Figures \ref{fig:smc} and \ref{fig:racs_gp} show objects with a range of angular scales up to $\sim$15 arcminutes.

Artefacts in the RACS images come in two varieties. The difficulty mentioned above with accurate use of the short baselines, in spite of excluding the very shortest, has led to broad artefacts positioned about bright extended sources, typically at a distance of 40 -- 60 arcminutes and having total flux-density of 10 -- 20 per cent of the parent source. The second variety is an elevated variation in the background noise close to strong compact sources. The second term in expression \ref{eqn:delta-s} is in part a reflection of this, but around some very strong sources the problem is more severe. There is strong evidence that this variety of artefact is less common in more recently observed fields.  Data collected earlier suffered from more system instability in beam bandpass shape.

\begin{figure*}
	\includegraphics[width=\textwidth]{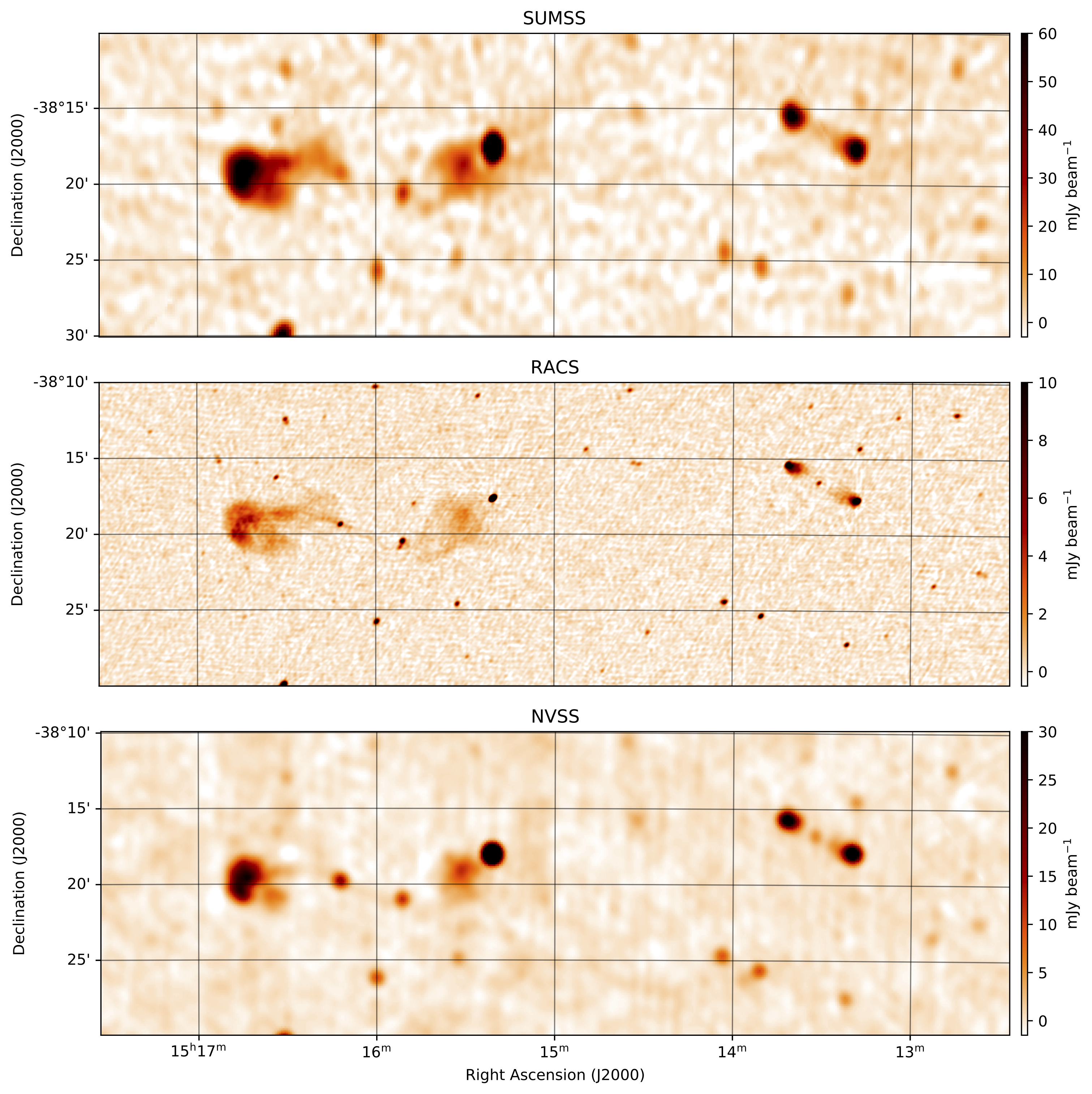}
    \caption{RACS, NVSS and SUMSS images of the same region. In the RACS image, the PSF has dimensions 16 $\times$ 11 arcseconds, and the image noise is 200\,$\mu$Jy/beam.}
    \label{fig:racs_nvss_sumss}
\end{figure*}

Figures \ref{fig:racs_gp} and \ref{fig:racs_nvss_sumss} provide some comparison between images from RACS and other radio surveys.  Figure \ref{fig:racs_gp} shows a portion of the Galactic Plane from the RACS and GLEAM surveys \citep{HurleyWalker:2016kq}. The GLEAM image has superior sensitivity to large angular scales, and the RACS image has the better resolution. Also evident is the differing spectral energy distribution of the objects in the field: thermal emission from H\,\textsc{ii}  regions is more prominent in the RACS image, while non-thermal emission from super-nova remnants is brighter in the lower frequency GLEAM image.

Figure \ref{fig:racs_nvss_sumss} shows a field imaged by all of NVSS, SUMSS and RACS.  The comparison shows clearly how better sensitivity and resolution of the RACS observations lead to more sources being detectable, and to better definition in the extended objects.

\subsubsection{Influence of Solar system objects}
Most of the first epoch of RACS observations were made within a three-week period (2019APR--MAY) and so inevitably Solar system objects appear in some images. Regions within about 6 degrees of the Sun itself were avoided and observed at a later date. A number of tiles close to the Sun bear the effects of contamination by solar radiation on the short baselines, which were minimised by excluding data from baselines shorter than 100\,metres.

The Moon appears in two of the tiles, which were also re-observed later.  Jupiter is visible close to the boundary of two tiles, observed on 2019APR25 and 2019APR27. It appears as an extended source of total flux-density $\sim$3.8~Jy. Saturn and Mercury are also visible with flux-densities of 20\,mJy and 4\,mJy respectively, although Mercury's rapid motion required it to be tracked in phase for a proper detection.

\section{RACS image release}
\label{sec:datarelease}
Here we have reported on the first RACS pass over the sky in the 743.5--1031.5\,MHz band. Results from this first stage of the survey are being released on CASDA as a set of data products for each of the 903 fields. Fields are named for their central J2000 position on the sky as \texttt{RACS\_hhmm$\pm$dd} where \texttt{hh}, \texttt{mm} and  \texttt{dd} are each the two digits for hours and minutes of Right Ascension and degrees of Declination, respectively. The data products released for each field are listed below.

\begin{enumerate}
    \item Total-intensity images for both the zeroth and first Taylor terms. These are approximately 7\,degrees in extent and have been brightness corrected as described in Section \ref{subsubsec:fluxscale}. The PSF size varies over these images so the estimation of total source flux-densities requires the procedure described in Section \ref{subsubsec:psf} to be followed.
    \item Images as for item (1) but processed to have a uniform PSF across the image.
    \item Weights images, needed for proper combination of adjacent tiles.
    \item Images of the image noise determined over 250-arcsecond wide cells, as introduced in Section \ref{subsubsec:noise}.
    \item Measurement sets holding calibrated visibility data, corrected for polarisation leakage at the beam centre. Note that although the visibility data for all four polarization products (XX, XY, YX, YY) are included, the formation of Stokes quantities (I, Q, U, V) requires knowledge of the orientation of the antennas' X and Y polarisation planes on the celestial sphere.
    \item Images of the effective PSF parameters over the field of view: major and minor axes of the fitted elliptical gaussian, and its position angle.  The brightness scale in each total-intensity image is correct; any software tool used to estimate the integrated flux-density of sources in the image will need the dimensions of the effective PSF.
\end{enumerate}
All images are presented in FITS format, and the visibility data stored in CASA-style MeasurementSets.

We also make available a database\footnote{RACS field data base at \url{https://bitbucket.csiro.au/projects/ASKAP_SURVEYS/repos/racs}} holding information about each field such as the circumstances of its observation and some average properties of the field image. These data are accompanied by software tools for accessing the database, and include procedures for performing the flux-density corrections mentioned above.

Additional items will be added to the CASDA release in future:
\begin{enumerate}
    \item Additional, lower quality images if the fields with multiple observations; these may be useful for variability studies.
    \item Images of 6.3\,degree extent, but with full sensitivity to the image boundaries. These will be prepared by linear combination with adjacent tiles.
\end{enumerate}

Observations for RACS will continue, specifically to sample other parts of the spectrum accessible to ASKAP. Images from those future observations will be added to the RACS archive on CASDA.

\section{SUMMARY}
\label{sec:summary}

We have introduced the Rapid ASKAP Continuum Survey and the publication of its first total-intensity images covering the whole sky south of declination $+41$\,degrees. In Section \ref{sec:survey_results} we have characterised the performance of ASKAP during the RACS observations and have presented the corresponding properties of the survey data products. These we summarise in Table \ref{tab:survey_properties}.

\begin{table*}
\caption{Survey properties}
\centering
\begin{tabular}{rcl}
\hline\hline
Number of images & & 903  \\
Image size & & 60  sq deg\\
Total survey area & & 34240  sq deg\\
Repeated fields & & 318  \\
Reference frequency & & 887.5  MHz\\
Bandwidth & & 288   MHz\\
Polarisation & & Stokes I  \\
Image type & &  MFS\textsuperscript{1}; Taylor terms  0,1 \\
Astrometric precision & $\Delta \theta$ & 0.8  arcsec\\
Astrometric accuracy & $\Delta \alpha \cos \delta$ & -0.6$\pm$0.6  arcsec \\
                     & $\Delta \delta$ &  -0.4$\pm$0.7  arcsec\\
Point-spread-function & $B_{min}$ & 9.8 < 11.8$\pm$0.9 < 15.5  arcsec \\
& $B_{maj}$ & 12.0 < 18.0$\pm$4.3 < 33.1  arcsec \\
Image noise & (tile medians)& 173 < 261$\pm$66 <  984 $\mu$Jy/beam \\

 \hline\hline
 \multicolumn{3}{l}{\textsuperscript{1}\footnotesize{MFS = multi-frequency synthesis}} \\

\end{tabular}
\label{tab:survey_properties}
\end{table*}

Over most of the surveyed area, the RACS images are of high quality.  The low spatial frequencies in bright extended sources have been difficult to represent well, partly for lack of rotational synthesis. Data quality improved through the course of the RACS observations (2019APR to 2020JUN); management of the phased-array feeds has improved, and improved scheduling has led to more compact image point-spread-functions.

The RACS dataset has enabled a thorough analysis of the ASKAP flux-density calibration and will inform future changes to standard telescope operating procedures. Beams shapes will be measured routinely, and the results used by mosaicing software to produce images with flat flux-density scale across their extent.

The images we release will provide a useful reference image of the radio sky in the 30\,cm band, bridging the spectrum between low-frequency surveys such as TGSS and GLEAM and NVSS at 1.4\,GHz.  Applications will include the search for radio transients. A total intensity source catalogue is being prepared from the RACS images (RACS paper II).

RACS observations will continue at the higher ASKAP frequencies, allowing the Global Sky Model to be applicable to all ASKAP surveys.


\begin{acknowledgements}
TM acknowledges the support of the Australian Research Council through grant FT150100099.
JP and JL are supported by Australian Government Research Training Program Scholarships.
The Australian Square Kilometre Array Pathfinder, Australia Telescope Compact Array, and Parkes Radio Telescope are part of the Australia Telescope National Facility which is managed by CSIRO. 
Operation of ASKAP is funded by the Australian Government with support from the National Collaborative Research Infrastructure Strategy. ASKAP uses the resources of the Pawsey Supercomputing Centre. Establishment of ASKAP, the Murchison Radio-astronomy Observatory and the Pawsey Supercomputing Centre are initiatives of the Australian Government, with support from the Government of Western Australia and the Science and Industry Endowment Fund. 
We acknowledge the Wajarri Yamatji as the traditional owners of the Murchison Radio-astronomy Observatory site. 

We thank the anonymous referee for careful reading and comments on the manuscript that have led to improvements in this paper.

\end{acknowledgements}

\begin{appendix}
\section{Computing the flux-density scale corrections}
\label{app:tt1corr}

Explanation of the brightness-scale correction procedure justifies a brief review of linear mosaicing, the determination of sky brightness as the linear combination of brightness measures from several overlapping images.  Let $B_t = B_t(l,m)$ be the true sky brightness at position $(l,m)$. We image the sky with $n$ beams $i = 0, .. n-1$ and in each we measure a brightness $b_i(l,m)$, from which we estimate the measured brightness $B_m(l,m)$ as
\begin{align}
\label{eqn:linmos}
    B_m &= \frac{\sum_i w_i \frac{b_i}{A_i}}{\sum w_i} = \frac{\Sigma b_i A_i}{\Sigma A_i^2}
\end{align}
where $A_i$ is the modelled amplitude of the $i$th beam at $(l,m)$ ($0 \leq A_i \leq 1$) and $w_i = w_i(l,m)$ is a direction-dependent weight.  We weight each image by its inverse variance and assume that $w_i(l,m) = (\sigma_0/A_i(l,m))^{-2}$ where $\sigma_0$ has the same value for all beams.  In practice this assumption is violated, but the errors introduced in this application are typically less than one percent. 

With knowledge of the true beam shapes $H_i$, we can write $b_i = B_t H_i$ so that, after rearrangement, an expression for the true brightness is
\begin{align}
    \label{eqn:fluxcorr_a}
    B_t &= B_m \frac{\sum A_i^2}{\sum c_i  H_i A_i}
\end{align}

The beams of the RACS footprint were measured using the standard holography technique \citep{Hotan:2016uk} on 2020FEB06. That measurement yielded values of $H$ for each beam on each antenna (excluding the antenna used as the holography reference) and as a function of frequency.  Asymmetries in beam shape, particularly evident in beams far from the field centre, lead to the point of maximum response falling closer to the optical axis than the nominal beam centre. Consequently the calibration source is not placed at the beam maximum, introducing an error in its inferred amplitude.  The additional beam-specific term $c_i$ included in equation \ref{eqn:fluxcorr_a} corrects for this effect, and is determined from  $H_i$ in the direction of the nominal beam centre.  

The expression \ref{eqn:fluxcorr_a} can be calculated as a function of frequency $\nu$: $f = B_t/B_m = f(\nu)$. This allows the determination of corrections to both the zeroth and first Taylor term images.  The intensity at the imaged position in the sky is written as 
\begin{align}
    \label{eqn:taylor}
    I_\nu &= I_0 + \frac{\nu - \nu_0}{\nu_0}I_1 + ...
\end{align}
where $I_0$ and $I_1$ are the zeroth and first Taylor-term images produced by the ASKAPsoft pipeline. Expanding $f(\nu)$ in the same way 
\begin{align}
    \label{eqn:f_expand}
    f_\nu &= f_0 + \frac{\nu - \nu_0}{\nu_0}f_1 + ...
\end{align}
allows the corrected intensity to be written as
\begin{align}
    \label{eqn:inufnu}
    I_\nu f_\nu &= [I_0 + \frac{\nu - \nu_0}{\nu_0}I_1][f_0 + \frac{\nu - \nu_0}{\nu_0}f_1] \nonumber \\
    &= I_0f_0 + \frac{\nu - \nu_0}{\nu_0}[I_1f_0 + I_0f_1] + ...
\end{align}
This was used to correct both the zeroth and first Taylor-term images.
\end{appendix}

\bibliographystyle{pasa-mnras}
\bibliography{racs}

\end{document}